\documentclass[%
superscriptaddress,
preprint
preprintnumbers,
 amsmath,amssymb,
 aps,
 showpacs,
 prX,
floatfix,
]{revtex4-1}

\usepackage[pdftex]{graphicx}% Include figure files
\usepackage{dcolumn}% Align table columns on decimal point
\usepackage{bm}% bold math
\usepackage{amsmath,amssymb,mathtools}
\usepackage[margin = 1cm, top = 1in,inner=1in,outer=1in]{geometry}
\usepackage{appendix}

\usepackage{url}
\graphicspath{{./figures/}}
\DeclareGraphicsExtensions{.pdf,.png,.jpg,.eps}

\def\dd{\mathrm{d}}
\renewcommand\d[1]{\textrm{d}#1}

\def\tsim{\sim}
\def\ggamma{\gamma}
\newcommand{\tr}{\textrm{Tr}}

\newcommand{\ddH}[2]{\frac{\delta^2 H}{\delta #1 \delta#2}}
\newcommand{\obs}[2]{\xi_{#1}^{(#2)}}
\newcommand{\obsj}{\xi_j^{(\alpha)}}
\newcommand{\Tr}{\mathrm{tr}}

\begin{document}

%\title{Simultaneous Bayesian reconstruction of diffusivity and bond
%  potentials using path integrals}

\title{Bayesian field theoretic reconstruction of bond potential and bond mobility in
single molecule force spectroscopy}

\author{Joshua C. Chang}\email{chang.1166@mbi.osu.edu}
\homepage{http://iamjoshchang.com}
\affiliation{Mathematical Biosciences Institute, The Ohio State University}
\author{Pak-Wing Fok}\email{pakwing@udel.edu}
\affiliation{Department of Mathematical Sciences, University of Delaware}
 \author{Tom Chou}\email{tomchou@ucla.edu}
 \affiliation{Departments of Biomathematics and Mathematics, University of California Los Angeles}

\begin{abstract}
Quantifying the forces between and within macromolecules is a
necessary first step in understanding the mechanics of molecular
structure, protein folding, and enzyme function and performance.  In
such macromolecular settings, dynamic single-molecule force
spectroscopy (DFS) has been used to distort bonds. The resulting
responses, in the form of rupture forces, work applied, and
trajectories of displacements, have been used to reconstruct bond
potentials. Such approaches often rely on simple parameterizations of
one-dimensional bond potentials, assumptions on equilibrium starting
states, and/or large amounts of trajectory data. Parametric approaches
typically fail at inferring complex-shaped bond potentials with
multiple minima, while piecewise estimation may not guarantee smooth
results with the appropriate behavior at large distances. 
Existing techniques, particularly those based on work theorems, also do not
address spatial variations in the diffusivity that may arise from
spatially inhomogeneous coupling to other degrees of freedom in the
macromolecule, thereby presenting an incomplete picture of the overall
bond dynamics.
% Finally, the inherent $10-20\%$ error in the calibration of the 
% device spring constant using existing methods contributes significantly to the uncertainty 
%of the resulting solution.
 To solve these challenges, we have developed a
comprehensive empirical Bayesian approach that incorporates data and
regularization terms directly into a path integral. All experiemental and statistical
parameters in our method are estimated empirically directly from the data.
%This method determines
%the device spring constant directly from observations of its Brownian motion.
 Upon testing our method on simulated data, our regularized approach requires fewer data
and allows simultaneous inference of \emph{both} complex bond
potentials \emph{and} diffusivity profiles.  Crucially, we show that
the accuracy of the reconstructed bond potential is sensitive to the
spatially varying diffusivity and accurate reconstruction can be
expected only when both are simultaneously inferred.  Moreover, after
providing a means for self-consistently choosing regularization
parameters from data, we derive posterior probability distributions,
allowing for uncertainty quantification.
\end{abstract}

\maketitle

\section{Introduction}

Inverse problems involving random walks are encountered throughout
the sciences. In these problems, one seeks to reconstruct one or more
functions that describe the dynamics of the random process, from
measurements of trajectories or first-exit times.  Examples
include the reconstruction of absorption and scattering profiles in
diffuse optical tomography~\cite{arridge1999optical} and inference of
stochastic volatility in
finance~\cite{coleman1998reconstructing,reno2008nonparametric}.  

Such inverse problems also arise in molecular biophysics, in which one
wishes to infer molecular energy
landscapes~\cite{evans1995sensitive,heymann2000dynamic,merkel1999energy,neuman2008single,lang2004simultaneous,hinterdorfer2006detection,rawicz2008elasticity,koch2003dynamic,jobst2013investigating,
  maitra2010model,hummer2001free,hummer2010free} relevant to protein
interactions~\cite{rief1997single,puchner2009force,fernandez2004force},
chromosome and DNA structure~\cite{dobrovolskaia2012dynamics,
  ros2004single,rief1999single,clausen2000force},
biorecognition~\cite{rief1997single,rief1999single,ros2004single}, and
cellular structure~\cite{helenius2008single,anselmetti2007analysis,
  benoit2000discrete,evans2007forces}.  In these applications, dynamic
force spectroscopy (DFS) is typically used to pull apart molecules or
bonds along one direction in a complex high-dimensional energy
landscape (see Fig.~\ref{fig:fig1}).  Much of the existing literature
on this inverse problem has focused on recovery of the underlying
molecular bond potential based on rupture force
statistics~\cite{dudko2008theory,dudko2009single,merkel1999energy,lang2004simultaneous,freund2009characterizing,fuhrmann2008refined,evstigneev2003dynamic}.
%
%\begin{figure}[h!]
%\centering
%\includegraphics[width=8.7cm]{fig2}
%\caption{\textbf{Effective potential.} (a)
%Shown in units of $k_{B}T$ are a representative underlying,
%  time-independent molecular potential $U(x)$ (solid black), the
%  harmonic potential of the pulling device (dashed black), and the
%  effective potential $\Phi(x,t) = U(x)+K(x-8)^{2}/2$. Here, the bond
%  coordinate $x$ is in arbitrary units and $K=0.3$. (b) The total
%  potential $\Phi(x,t)$ for $L(t)=Vt=3,5,7,9,11$. Note that at
%  intermediate times (pulled distances) the total potential can
%  develop a double-well structure provided $K$ is sufficiently small.
%  (c) $U(x)$, $K(x-8)^{2}/2$, and $\Phi(x,t)$ for a stiffer pulling
%  device with $K=1$. (d) $\Phi(x,t)$ for $Vt=3,5,7,9,11$. Note that
%  $\Phi(x,t)$ never develops a double-well structure in this case.
%\label{fig:fig2}}
%\end{figure}
% 
While such approaches allow reconstruction of simple parametric forms
of the bond potential, they require careful tuning of experimental
parameters. For example, the pulling device cannot be too stiff if a
transient barrier and rupturing behavior is desired
~\cite{shapiro1997quantitative}. Moreover, event-based reconstruction
requires pulling over a range of carefully tuned speeds.  Most
importantly, reconstruction based on rupture forces also ignores the
full wealth of information contained in measurements of the individual
displacements.

% which would allow more direct inference of spatial
%structure of the bond force.

Indeed, there exists extensive literature on drift recovery for random
walks using trajectory measurements and/or work theorems
\cite{hummer2001free,hummer2005free,hummer2010free,balsera1997reconstructing,block_review}.  These
approaches typically involve discretization of the solution
domain~\cite{turkcan2012bayesian,masson2014mapping,schuss2011nonlinear},
where piecewise-constant solutions are obtained through binwise
Bayesian inference, maximum likelihood, or moment-matching as in the
case of work theorems~\cite{hummer2010free,alemany2012experimental,seifert2012stochastic}. Not only do these
approaches require sufficient sampling of distributions of
displacements or work, but they also cannot be easily adapted to simultaneous
reconstructions of functions such as diffusivity.  

In fact, the diffusivity cannot be independently extracted using work
theorem-based reconstructions.  However, spatial variations in diffusivity are
intertwined with displacement trajectory-based recovery of the
underlying bond potential. Variations in diffusivity are
associated with varying landscape ``roughness'' \cite{zwanzig1988},
which ultimately arises from projections of higher-dimensional
trajectories onto the path defined by the external pulling
\cite{best2010coordinate}. Thus, spatially varying diffusivity
contains information on how a high-dimensional system projects down to
form a one-dimensional potential profile.

%
%SOMETHING ABOUT PAST WORK ON PIECE-WISE CONSTANT APPROACHES (SCHULTEN)
%We also develop an efficient numerical scheme and
%apply our proposed procedure to simulated data.
%
%Inference schemes based on trajectories are also effectively simpler
%than inference based on inverting the two-dimensional (space-time)
%solution to an associated Fokker-Planck equation. Since the objects of
%recovery are time-independent, the time-dependent portion of the
%drift can be subtracted, thereby reducing the dimensionality of the
%problem.  Furthermore, trajectory-based inference can be based on
%principles like \emph{Maximum Likelihood}, which come with theoretical
%guarantees on asymptotic efficiency.
%
Regardless of inversion method, samples of Brownian trajectories are
taken pointwise, meaning that the recovery of continuous functions
governing Brownian motion is ill-posed.  Since numerical inversion of
the drift or diffusion functions will be at best ill-conditioned
\cite{fok2010reconstruction}, inference on random walks is typically
performed at a certain spatial resolution wherein averaging of
observations
occurs~\cite{masson2014mapping,turkcan2012bayesian,schuss2009theory,schuss2011nonlinear}.
However, this type of procedure does not guarantee stability or
smoothness of the reconstructed functions.

Recently, Bayesian path integral-based approaches have been developed
for the recovery of mathematically continuous solutions, where
candidate reconstructions are weighted by properties encoded in a
distribution that reflects \emph{a priori} knowledge. In this vein,
Lemm, Uhlig, and Weiguny \cite{lemm2000bayesian} demonstrated such an
approach for the recovery of potential functions from paths observed
in quantum systems. We will show that using this type of approach in
the DFS setting naturally incorporates the simultaneous reconstruction
of both diffusivity and bond potential.  Bayesian theory then provides
a procedure for inference, uncertainty quantification, and parameter
identification. The application of Bayesian theory in this way also
defines the inverse problem in its more-natural continuum
representation using partial differential equations (PDEs).  Any
discretization used in solving the PDEs is independent of the problem
formulation.

%
%It is important to take such
%considerations in effect when for instance determining the effect of
%temperature on the bond force as diffusivity goes linearly with
%temperature. This fact means that the drift scales linearly with
%temperature while the noise term $\sqrt{D}$ goes as $\sqrt{T}$.
%
%The influence of force and diffusivity on the motion of the bond
%coordinate is local. For this reason, one needs to observe local
%movement in order to reconstruct these functions. Due to the discrete
%nature of sampling, one never truly observes a trajectory passing
%through any particular location.
%
Here, we develop a path integral-based empirical Bayesian procedure to
reconstruct both bond forces \emph{and} diffusivities directly from
trajectory measurements. Our method is general in that we need make no
assumption about the pulling protocol or device spring constant; the
only assumption made is applicability of the Brownian motion. We
provide an efficient numerical procedure, test our approach on
simulated trajectories, and show that very reasonable numbers of
trajectories are sufficient to simultaneously reconstruct complex
multi-minima bond potentials and diffusivities.  The sensitivity of
bond-force reconstruction to the diffusivity profile is also explored
and a physical interpretation of our regularization discussed.

\section{Methods}

\begin{figure}
\includegraphics[width=8.6cm]{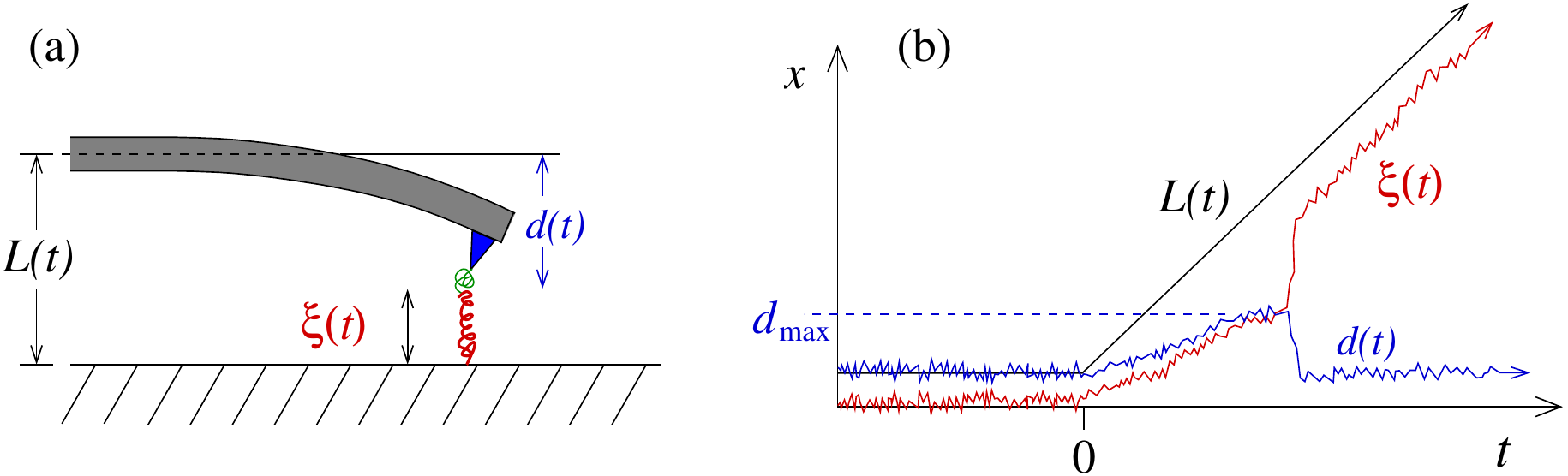}
\caption{\textbf{Dynamic Force Spectroscopy (DFS) setup and
    measurement.}  (a) Schematic of a DFS pulling experiment.  A
  pulling device with spring constant $K$ and reference control
  position $L(t)$ is attached to one end of a bond. As the device is
  lifted, it deflects by amount $d$, but also stretches the observed bond
  coordinate $\xi$, which is a mesurement of the underlying true bond coordinate $X$. 
  (b) Schematic of trajectories for $L(t)$, $d(t)$, and 
$\xi(t)\equiv L(t)-d(t)$. In reconstructions based on rupture
  forces, the maximum value $d_{\rm max}$ determines the force at
  rupture, indicated by the sharp increase in
  $\xi(t)$. \label{fig:fig1}}
\end{figure}

%%%%%%%%%%%%%%%%%%%%%%%%%%%%%%%%%
\vspace{2mm}%%%%%%%%%%%%%%%%%%%%%
%%%%%%%%%%%%%%%%%%%%%%%%%%%%%%%%%
\noindent \textbf{Problem set-up} Figure~\ref{fig:fig1} shows a
schematic of DFS in which a bond is pulled apart along the spatial
direction $x$, while the bond displacement $\xi(t)$ is measured and
recorded.  We assume that the bond coordinate is an over-damped random
variable and that is well-described by a stochastic differential
equation of the form

\begin{equation} \dd \xi = A(\xi,t)\dd t + \sqrt{2D(\xi)}\dd W,
\label{eq:sde}
\end{equation}
where $W$ is a Wiener white noise process, $D(x)$ is the
space-dependent diffusivity function, and $A(x,t)$ is the spatially
varying drift. Interpreting Eq.~\ref{eq:sde} using It\^{o} calculus we
find that the drift takes the form $A(x,t)\equiv -D(x)\partial_{x}
\Phi(x,t)+\partial_{x}D(x)$, where $\Phi(x,t)$ is the total potential.
The motion described by this drift term results from forces arising
from a potential gradient and a diffusivity gradient.  This definition
of $A(x,t)$ yields the expected Fokker-Planck equation (FPE) for the
probability distribution function $P(x,t)$: $\dot{P}(x,t) -
\partial_{x}(P D(x)\partial_{x}\Phi) =
\partial_{x}(D(x)\partial_{x}P)$ \cite{seifert2012stochastic}.

The total dimensionless (normalized by $k_{\rm B}T$) potential
$\Phi(x,t)$ is composed of the molecular bond potential $U(x)$ and a
moving harmonic potential arising from the pulling device (typically
an optical trap or AFM cantilever, as shown in
Fig.~\ref{fig:fig1}). The origin $L(t)$ of the harmonic potential is
controlled by the pulling device. Together, the effective potential
takes the form

\begin{equation}\label{eq:phi}
\Phi(x,t) = \underbrace{U(x)}_{\textrm{bond}} +
\underbrace{\frac{K}{2}(x-L(t))^2}_{\textrm{harmonic}},
\end{equation}
where $K$ is the device spring constant.
After differentiating Eq.~\ref{eq:phi} one finds

\begin{equation}
A(x,t) = D(x)\big[F(x)+\overbrace{K(L(t)-x)}^{F_a}\big] + D^\prime(x)\label{eq:drift}
\end{equation}
where $F(x)=-\dd U(x)/\dd x$ is the intermolecular bond force, and $F_a$ is the 
force applied by the pulling apparatus.  In practice, the pulling device is moved at a constant velocity $V$ starting from an initial position $L_{0}$: $L(t) =
L_{0}+Vt$. Eq.~\ref{eq:drift} shows that pulling (increasing $L(t)$)
increases the drift thereby encouraging displacement of the bond
coordinate away from $x=L_0$. The goal of such experiments is to infer
properties of the bond potential $U(x)$, from how the bond coordinate
responds as $L(t)$ is increased.

 The bond force $F(x)$ will be
assumed to be a smooth continuous function that 
will be decomposed in the form

\begin{equation}
F(x) = F_{\rm d}(x)+ f(x),\label{eq:bondforce}
\end{equation}
where $F_{\rm d}(x)=\kappa x^{-\nu}$ ($\kappa\geq 0, \nu>1$) is the
most divergent component of the force associated with the divergent
part of the potential $U(x)\tsim x^{-\nu}$ ($\nu > 1$) as $x\to 0$. At
large separations, we assume the total force vanishes and
$f(x\to\infty)\to 0$. The behavior of $F$ near $x=0$ is not
particularly interesting, so we will make the simplifying assumption
that $F_{\rm d}(x)=6(x/2)^{-7}$, and restrict our recovery problem to
the region $[L_0,\infty)$. Ultimately, our reconstruction for the
potential and diffusivity for $x > L_{0}$ will not be too sensitive to
the exact form of the divergence; there will be very few trajectories
that sample the strongly repulsive region where $x$ is small.  The
smooth function $f(x)$ captures all other features of the
intermolecular bond force we wish to reconstruct.  We impose vanishing
boundary conditions at $x=0$ and $x\to\infty$, but do not assume
$f(x)$ obeys any particular parametric form. In our subsequent inverse
problem, since $F_{\rm d}(x)$ is specified, and molecular forces are
conservative, the reconstruction of $f(x)$ will be equivalent to
reconstruction of $F(x)$ and, up to an additive constant, the
molecular potential $U(x)$.

%%%%%%%%%%%%%%%%%%%%%%%%%%%%%%%%%
\vspace{2mm}%%%%%%%%%%%%%%%%%%%%%
%%%%%%%%%%%%%%%%%%%%%%%%%%%%%%%%%
\noindent {\bf Empirical Bayes formulation} - Since the recovery of
continuous $f(x)$ directly from discrete data is ill-posed, we now
describe a path integral-based Bayesian interpretation of the
so-called Tikhonov
regularization~\cite{chang2014path,lemm2000bayesian,ensslin2009information,cotter2009bayesian,heuett2012bayesian,farmer2007bayesian,stuart2010inverse}.
The key feature this method is the usage of a smoothness penalty to
select solutions from particular well-behaved function spaces. The
choice of function space and smoothing is considered \emph{prior
  knowledge} and is determined either from physical considerations or
estimated directly from the data.  The inverse problem is then
investigated through the evaluation of a partition function, using a
path integral over the given function space. A general form of
Tikhonov regularization manifests itself through a prior probability
density on $f(x)$ of the form

\begin{equation}
\pi(f|\boldsymbol{\theta}) = \mathcal{Z}^{-1}_f \exp\left\{-\frac{1}{2}\int_0^\infty
f(y)R_f(-\Delta)f(y) \dd y\right\},\label{eq:Mf}
\end{equation}
where $\Delta$ is the Laplacian operator,
$R_f$ is a self-adjoint pseudo-differential regularization
operator containing some parameters $\boldsymbol{\theta}$, 
 and $\mathcal{Z}_f$
is a normalization factor.  We assume for now that we know $R_{f,g}$ and their associated parameters $\boldsymbol\theta$.
A more thorough discussion on their choice is presented in the next section.

 To enforce the positivity of $D(x)$, we
express diffusivity in terms of the log-diffusivity

\begin{equation}
g(y) = \log\frac{D(y)}{D_0}\label{eq:gy}
\end{equation}
where $D_0>0$, a uniform background diffusivity, can be estimated
directly from the data (see Eq.~\ref{eq:D0} given in the \textsc{Supplemental Methods}). We
assume a similar prior distribution on the log-diffusivity $g(y)$ of
the form

\begin{equation}
\pi(g|\boldsymbol{\theta}) =\mathcal{Z}^{-1}_g \exp\left\{-\frac{1}{2}\int_0^\infty g(y)R_g(-\Delta)g(y)\dd y\right\}.\label{eq:Mg}
\end{equation}
The normalization factors $\mathcal{Z}_f, \mathcal{Z}_g$ do not affect
the inference of $f(x)$ and $g(x)$, but are important when one wishes
to self-consistently determine a specific form of regularization $R_{f,g}$.
Eqs.~\ref{eq:Mf} and \ref{eq:Mg} enforce that the prior probability
distributions are over a collection of functions $f(x)$ and $g(x)$
that have Gaussian spatial auto-correlations.  These auto-correlations are
determined by the Green's functions of the
pseudo-differential-operators $R_f$ and $R_g$, which can be thought of
as kernels encoding certain magnitude and scale information about the
spatial variability in the set of functions $f$ and $g$.

Experimentally, a trajectory is composed of measurements of bond
displacements, $\boldsymbol{\xi} \equiv (\xi_1,\xi_2,\ldots,\xi_N)$,
taken at times $t_1,t_2,\ldots t_N$.
If the force $F(x)=F_d(x)+f(x)$ and diffusivity $D(x)=D_0e^{g(x)}$ are given, the
\emph{likelihood} or probability of observing a given trajectory
$\xi_{j}$ ($0\leq j \leq N)$ can be formulated in terms of the product
of transition probabilities
$\pi(\boldsymbol{\xi}|f,g)=\prod_j\Pr(\xi_{j+1}|\xi_j,f,g)$. In the
limit as $\delta t\to0$, the transition probabilities, interpreted
using It\^{o} rules, are themselves Gaussian with mean
$A(\xi_j,t_j)\delta t$ and variance $2D(\xi_j)\delta t$ (see Supplemental Eq.~\ref{eq:likelihood}
and the \textsc{Supplemental Methods} for the derivation).
 We have assumed that measurement times $t_{i}$ and 
displacements $\xi_{i}$ are precisely measured (the error remains small relative to $2D\delta t$), and that the sampling
frequency is sufficiently high ($\delta t = t_{j+1}-t_j$ is small).

Given a collection of $M$ independently measured trajectories
$\Xi=\{\boldsymbol{\xi}^{(\alpha)}\}, (1\leq \alpha \leq M)$, one can
write the total likelihood function for observing the entire
collection of trajectories as a product of the likelihoods of the
individual trajectories, 
\begin{align}
\lefteqn{\pi(\Xi|f,g) =  \prod_\alpha \pi(\boldsymbol{\xi}^{(\alpha)} | f,g)}\nonumber\\
&\quad= \exp\Bigg\{-\sum_{j,\alpha}\Bigg[ \frac{(\xi^{(\alpha)}_{j+1}-\xi^{(\alpha)}_j-A(\xi^{(\alpha)}_j,t_j)\delta t)^2}{4 D(\xi^{(\alpha)}_j)\delta t} \Bigg] \Bigg\} \nonumber\\
&\qquad\times\prod_{j,\alpha} \sqrt{\frac{1}{4\pi D(\xi^{(\alpha)}_j)\delta t}} \label{eq:likelihood1}.
\end{align}

Using Bayes rule, the \emph{posterior probability
distribution} for $f$ and $g$, given observation of $\Xi$ and regularization parameters
$\boldsymbol{\theta}$ is
\begin{equation}
\pi(f,g|\Xi,\boldsymbol{\theta}) ={\pi(\Xi|f,g)\pi(f|\boldsymbol{\theta})\pi(g|\boldsymbol{\theta})\over \pi(\Xi)}
\equiv {e^{-H[f,g|\Xi,\boldsymbol{\theta}]}\over \mathcal{Z}},
\end{equation}
where $\mathcal{Z}$ is a dimensionless normalization
constant and $H$ is an \emph{information Hamiltonian} given by

{\begin{align}
\lefteqn{H\left[f,g\mid\Xi,\boldsymbol{\theta} \right] = }\nonumber\\
&\quad\frac{1}{2}\int_0^\infty f(y)R_f(-\Delta)f(y)d y
%+\frac{1}{2}\tr\log R_f(-\Delta)
+\frac{1}{2}\int_0^\infty g(y)R_g(-\Delta)g(y)\dd y\nonumber\\
%+\frac{1}{2}\tr\log R_g(-\Delta) 
&+{\frac{1}{2}\sum_{\alpha,j}\log D(\xi^{(\alpha)}_j) + \sum_{\alpha,j}
\frac{{\left(\xi^{(\alpha)}_{j+1}-\xi^{(\alpha)}_j-A(\xi^{(\alpha)}_j,t_j)\delta
    t\right)^2} }{4D(\xi^{(\alpha)}_j)\delta t}},\label{eq:hamiltonian}
\end{align}}where the last two terms arise from taking the logarithm of the likelihood
given in Eq.~\ref{eq:likelihood1}. As a reminder, we have assumed that measurement noise is
neglible relative to the inherent stochastic noise of the Brownian motion at time scale $\delta t$.
Relaxation of this assumption would require the evaluation of an additional path-integral in $\xi$, as performed in
\citet{masson2014mapping,masson2009inferring}.

The most-probable reconstructions for $f(x)$, $g(x)$, minimize Eq.~\ref{eq:hamiltonian}.  These
reconstructions constitute the \emph{maximum-a-posterior solution}, or
the specific choice of force $F(x)=F_{\rm d}(x)+f(x)$ and diffusivity
$D(x)=D_0e^{g(x)}$ that minimizes Eq.~\ref{eq:hamiltonian}. They are
found by solving the coupled system of Euler-Lagrange equations

\begin{equation}
\frac{\delta H}{\delta f}=0 \,\,\,\mbox{and}\,\,\, \frac{\delta H}{\delta g} =0,
\label{eq:EL}
\end{equation}
%
%\label{eq:dHf}\\
%\frac{\delta H}{\delta g} &=0.\label{eq:dHg}
%\frac{\delta H}{\delta x^{(\alpha)}} & = 0%\\
%\nabla_{\boldsymbol{\theta}} H &= 0
%
and constitute the \emph{mean-field} or \emph{classical} solution. The
main difficulty in solving these equations lie in inverting a large
matrix of rank equal to the number of observed trajectory positions.
A computational method for approximating the solution about
interpolation points is presented in the \textsc{Supplemental Methods}. In this method,
``sufficient statistics'' of the data are computed only a single time, after which
optimization occurs in a lower-dimensional space. Furthermore, the sufficient statistics
are independent of the regularization parameters, allowing an arbitrary number of candidate solutions to
be computed without re-processing the data.

%%%%%%%%%%%%%%%%%%%%%%%%%%%%%%%%%
\vspace{2mm}%%%%%%%%%%%%%%%%%%%%%
%%%%%%%%%%%%%%%%%%%%%%%%%%%%%%%%%
\noindent
{\bf Regularization parameters and uncertainty quantification} - Up to
this point, we have assumed that one knows what to use for the
operators $R_f(-\Delta)$ and $R_g(-\Delta)$. Since these operators can
be thought of as prior information, their choice can be motivated from
physical considerations whenever such information is
available~\cite{chang2014path}. Typically, the uncertainty in the
reconstructed functions arise from the mathematical ill-posedness of
the inverse problem.  However, in the DFS problem, the 1D bond
potential is a projection from a high-dimensional macromolecular
stochastic process and the effective bond potential will suffer
physical thermal fluctuations that also contribute to its uncertainty.
Therefore, it is desirable to choose $R_{f,g}$ directly from the data, 
which may shed light on how ``orthogonal'' modes are thermally coupled to to
the 1D bond potential.

Note that if $R_{f,g}(-\Delta)=1$ is chosen as the regularization operator,
the corresponding Green's function is the Dirac $\delta-$distribution.
This situation corresponds to the spatially unregularized inverse
problem. Numerically, if this inverse problem is solved over a
discrete lattice, then solution is the recovery of piecewise constant
force and diffusivity. For a more physically realistic and better-behaved
 inversion, it is convenient to restrict $R_{f,g}(-\Delta)$ to a
family of operators that impose spatial regularity.
%
%A particularly suitable family in the context
%of Tikhonov regularization is the Bessel potential family of operators,
%\begin{equation*}
%M_s(-\Delta)=\beta\left(1-\frac{\Delta}{s\gamma}\right)^s
%\end{equation*}
%where $\gamma$ is a spatial-scale parameter, $s$ is  parameter that controls the degree of smoothness,
%and $\beta$ acts like a reciprocal temperature. The Bessel family of
%operators is associated with the Sobolev's spaces of functions $H^s$, which roughly stated,
%includes functions whose first $s$ derivatives are square-integrable.  These operators are
%also self-adjoint, so one can use them to form the separable Hilbert spaces
%$\{ \phi ,(\cdot,\cdot)| (\phi,\phi)<\infty\}$ where the inner product is defined
%\begin{equation*}
%(\phi,\varphi) = \beta\int \phi \left(1-\frac{\Delta}{s\gamma}\right)^s \varphi dx.
%\end{equation*}
%The free-space Greens functions for Bessel potentials are the Matern
%covariance kernels
%\begin{equation*}
%G_s(x,y) = \frac{1}{\beta}\frac{\pi^{1/2} 2^{1-s-1/2}}{\Gamma(s)\gamma^{s-1/2}}\left( \sqrt{\gamma}|x|\right)^{s-1/2}\mathcal{K}_{s-1/2}\left( \sqrt{\gamma}|x|\right)
%\end{equation*}
%where $\mathcal{K}_{(\ast)}$ is the modified Bessel function of the second kind.
%In the limit as $s\to\infty$, one obtains the heat flow operator
%
Henceforth, we will assume $f$ and $g$ are infinitely-differentiable and
use operators of the form
\begin{equation}
R_{f}(-\Delta)=\frac{e^{-{\gamma_{f}}\Delta/{2}}}{\beta_{f}
\sqrt{2\pi\gamma_{f}}}, \qquad R_{g}(-\Delta)=\frac{e^{-{\gamma_{g}}\Delta/{2}}}{\beta_{g}
\sqrt{2\pi\gamma_{g}}}.\label{eq:heatpdo}
\end{equation}
Using the operators in Eq.~\ref{eq:heatpdo}, one need only determine
two parameters for each field: the spatial scale $\gamma$ and the
inverse temperature $\beta$.  Assuming that no information is known
about these parameters, one may utilize any number of available information
theory-based methods, such as Bayesian model comparison or
maximum marginal likelihood (Empirical Bayes).  Here, we describe the
application of approximate maximum marginal likelihood to the problem
of choosing regularization parameters.

As its name implies, maximum marginal likelihood estimation seeks to
determine unknown parameters
$\boldsymbol\theta=(\beta_f,\beta_g,\gamma_f,\gamma_g)$ by maximizing
the marginal likelihood function
\begin{equation}
\pi(\Xi|\boldsymbol\theta)=\iint \mathcal{D}f\mathcal{D}g\,
\pi(\Xi|f,g)\pi(f |\boldsymbol\theta)\pi(g | \boldsymbol\theta)
\label{eq:marginal}
%\pi(\boldsymbol\theta).
\end{equation}
with respect to $\boldsymbol\theta$. This expression can be
interpreted as the probability of obtaining the observed data given
the regularization parameters $\boldsymbol\theta$.
The optimization of this quantity requires the evaluation of the path
integrals with respect to both fields $f$ and $g$. These integrals can
be approximated using the semiclassical
approximation~\cite{chang2014path} in which the Hamiltonian (Eq.~\ref{eq:hamiltonian})
is expanded about its extremal points $f^\star,g^\star$ to quadratic order
\begin{align}
H[f,g|\Xi,\boldsymbol{\theta}] \approx H[f^\star,g^\star|\Xi,\boldsymbol{\theta}]+ \frac{1}{2}
\iint \boldsymbol{\varphi}(y)^t \boldsymbol{\Sigma}^{-1}
\boldsymbol{\varphi}(z) \dd y \dd z.
\label{eq:semiclassicalH}
\end{align}
The difference of the functions from their classical solution is defined by
the new field
\begin{equation*}
\boldsymbol{\varphi}(x) =\left[\begin{matrix} f(x)-f^\star(x)\\
g(x)-g^\star(x)\end{matrix}\right],
\end{equation*}
and the semiclassical Hessian $\boldsymbol\Sigma^{-1}$ matrix is

\begin{equation}
 \boldsymbol{\Sigma}^{-1} = \left[\begin{matrix}
\frac{\delta^2 H}{\delta f(y)\delta f(z)} & \frac{\delta^2 H}{\delta f(y)\delta g(z)} \\
\frac{\delta^2 H}{\delta g(y)\delta f(z)} & \frac{\delta^2 H}{\delta g(y)\delta g(z)}
\end{matrix}\right]_{f^\star,g^\star}.
\label{eq:sigmainv}
\end{equation}
The probability distribution over the functions $f(x)$ and $g(x)$
has a spread defined by $\boldsymbol\Sigma$, which encodes the distribution of $f(x)$ and
$g(x)$ about their most likely values $f^*(x)$ and $g^*(x)$, thereby
providing an estimate of the errors in the estimates $f^*(x)$ and
$g^*(x)$.  Performing the resulting Gaussian path integral
$\mathcal{Z}^{-1}_f\mathcal{Z}^{-1}_g\int
\mathcal{D}{\boldsymbol\varphi}e^{-H[\boldsymbol{\varphi}\vert
    \Xi,\boldsymbol{\theta}]}$ yields the semiclassical approximation
to the negative of the marginal likelihood function
\begin{align}
\lefteqn{-\log\pi(\Xi|\boldsymbol\theta) = \textrm{const}+ H[f^{\star},g^{\star}\vert \Xi,\boldsymbol{\theta}]}\nonumber\\
&\quad + \tr \log \boldsymbol{\Sigma} - \tr\log G_{f}(x,y) -
\tr\log G_{g}(x,y)\label{eq:logmarginal},
\end{align}
where the additive constant is independent of the regularization
parameters and the $\tr\log G_{f}$ and $\tr\log G_{g}$ terms come from
the normalization terms $\mathcal{Z}_f$ and $\mathcal{Z}_g$.  Note
that an implicit $\boldsymbol{\theta}-$dependence arises in all terms
involving $R_{f,g}$, and the data-derived $f^{\star}$ and $g^{\star}$.
In the \textsc{Supplemental Methods}, we show that the computation of  Eq.~\ref{eq:logmarginal} 
is equivalent to the computation of the eigenvalues of a finite-dimensional matrix -- alowing for quick
evaluation of Eq.~\ref{eq:logmarginal} for use in standard optimization routines.

%%%%%%%%%%%%%%%%%%%%%%%%%%%%%%%%%%%%%%%%%%%%%%%%%%%%%%%%
\vspace{2mm}%%%%%%%%%%%%%%%%%%%%%%%%%%%%%%%%%%%%%%%%%%%%
%%%%%%%%%%%%%%%%%%%%%%%%%%%%%%%%%%%%%%%%%%%%%%%%%%%%%%%%

\noindent {\bf Reconstruction Procedure} - Summarizing, our general
procedure for simultaneous force and diffusivity reconstruction is:

\begin{enumerate}

\item If unknown, estimate the background diffusivity $D_0$ and the spring constant $K$ directly from data
using Supplemental Eqs.~\ref{eq:Kstar}-\ref{eq:D0}.

%\item Discretize the spatial domain of the problem by choosing a grid
%over which to solve the Euler--Lagrange equations.

\item For each choice of regularization parameters $\beta_{f,g}, \ggamma_{f,g}$:

\begin{enumerate}
\item Solve for the maximum {\it a posteriori} solution $f^\star,g^\star$
by solving Eqs.~\ref{eq:EL} using the
method outlined in the \textsc{Supplemental Methods}.

\item Compute the semiclassical variance matrix $\boldsymbol\Sigma$
by inverting the matrix in Eq.~\ref{eq:sigmainv}.

\item Compute the negative log-marginal likelihood given by Eq.~\ref{eq:logmarginal}
\end{enumerate}

\item Choose regularization parameters that minimize Eq.~\ref{eq:logmarginal}.

\end{enumerate}

\section{Results}

To demonstrate our method, we first simulated data from DFS pulling
experiments using two different bond potentials and diffusivities.
\begin{figure}
\includegraphics[width=8.6cm]{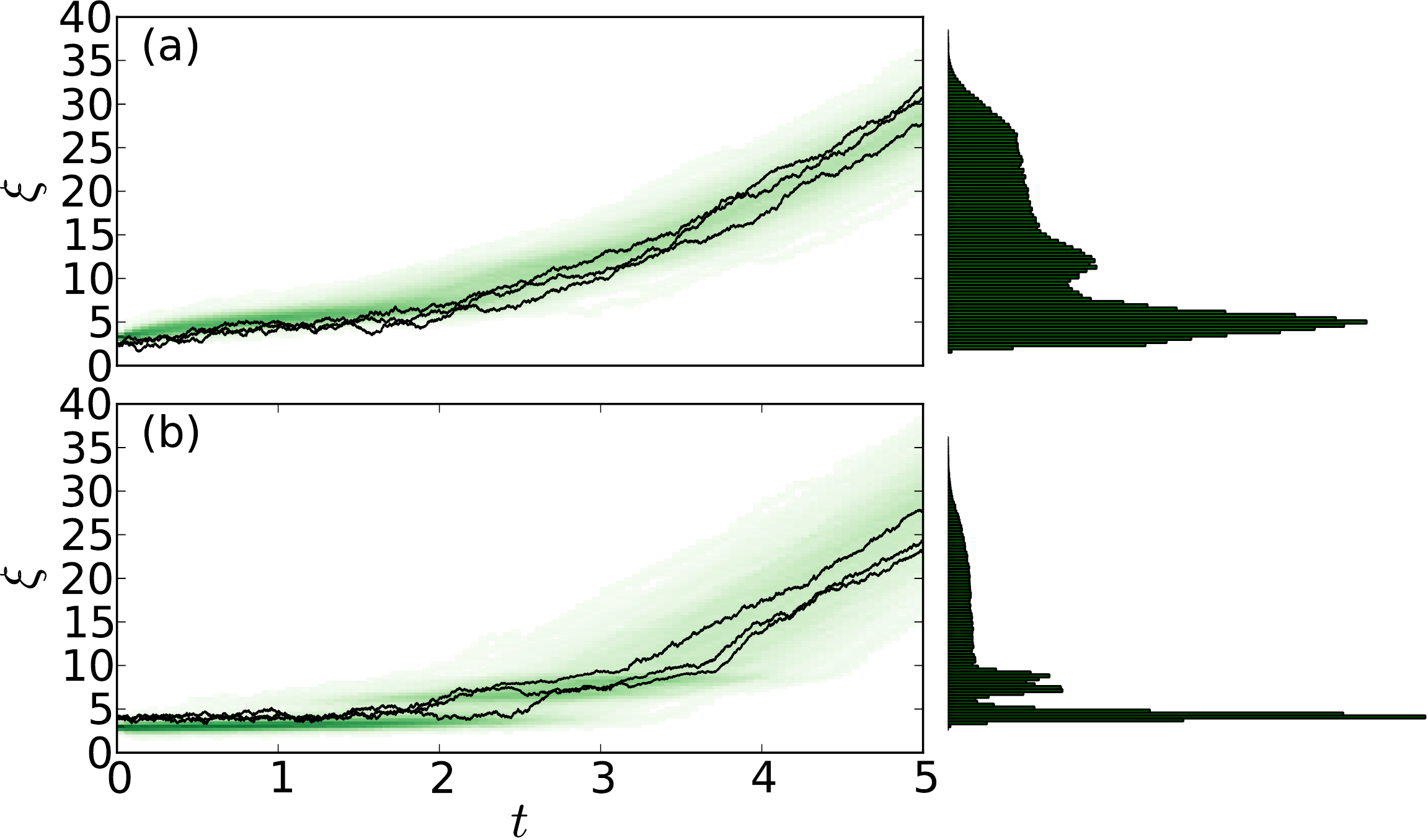}\centering
\vspace{-2mm}
\caption{\textbf{Trajectory data.}  Simulations using bond force and diffusivity
given by (a) Supplemental Eqs.~\ref{eq:D1},~\ref{eq:f1}, and (b) Supplemental Eqs.~\ref{eq:f2}.
  Three individual simulated
  trajectories (out of $10^{3}$) are shown in black.  Each trajectory
  represented a different pulling experiment of duration $5s$, sampled
  at $10\textrm{kHz}$, with $V=20,K=0.15$. The shaded region is
  compactly supported and represents the intensity of all $10^{3}$
  trajectories through each space-time point. While these trajectories
  are rather featureless, the histogram of positions observed across
  all trajectories (up to time 5s) is shown on the right and contains
  more features.  Each point in the histogram represents a single
  instance in which a position is sampled. Thus, each trajectory can
  sample a specific position many times. The total number of sample
  points is $10^{3}$ trajectories $\times$ $10$kHz $\times 5\mathrm{s}
  = 5\times 10^{7}$. These data can be aggregated across different
  experimental conditions and contain sufficient information with
  which to simultaneously reconstruct $f(x)$ and
  $g(x)$.\label{fig:fig2} }
\end{figure}
Fig.~\ref{fig:fig2} shows representative examples of simulated
trajectories.  Although the dynamics are governed by complex bond
potentials and spatially varying diffusivities, individual trajectories
are rather featureless.  The distributions that solve the associated
FPE are also qualitatively generic and featureless. However, data
across multiple trajectories can be aggregated as shown on the right
of Fig.~\ref{fig:fig2}.  

%Here, each point in the histogram is an
%instance in which a position is sampled. The total number of data
%points in this case is $10^{3}$ trajectories $\times$ $10$kHz $\times
%5\textrm{s} = 5\times 10^{7}$. These data contain the information with
%which potential and diffusivity reconstruction is possible.

Next, discrete measurements were extracted from our simulated
trajectories and used within our inference scheme in order to recover
the bond force and diffusivities that were used to generate the
simulated data in the first place.  We implemented our inference
method in Python 2.7.5 using the SciPy 0.14.0 library for numerical
optimization. The source code for our implementation is publicly
available at \url{https://github.com/joshchang/dfsinference}. In all
of the following examples, functions were recovered within the
interval from about $x=4$ to $x=32$, where $L_0=4$ was assumed to be
the starting point for the bond coordinate. In this interval, 200
evenly spaced interpolation points were chosen.

\begin{figure*}
\includegraphics[width=12.5cm]{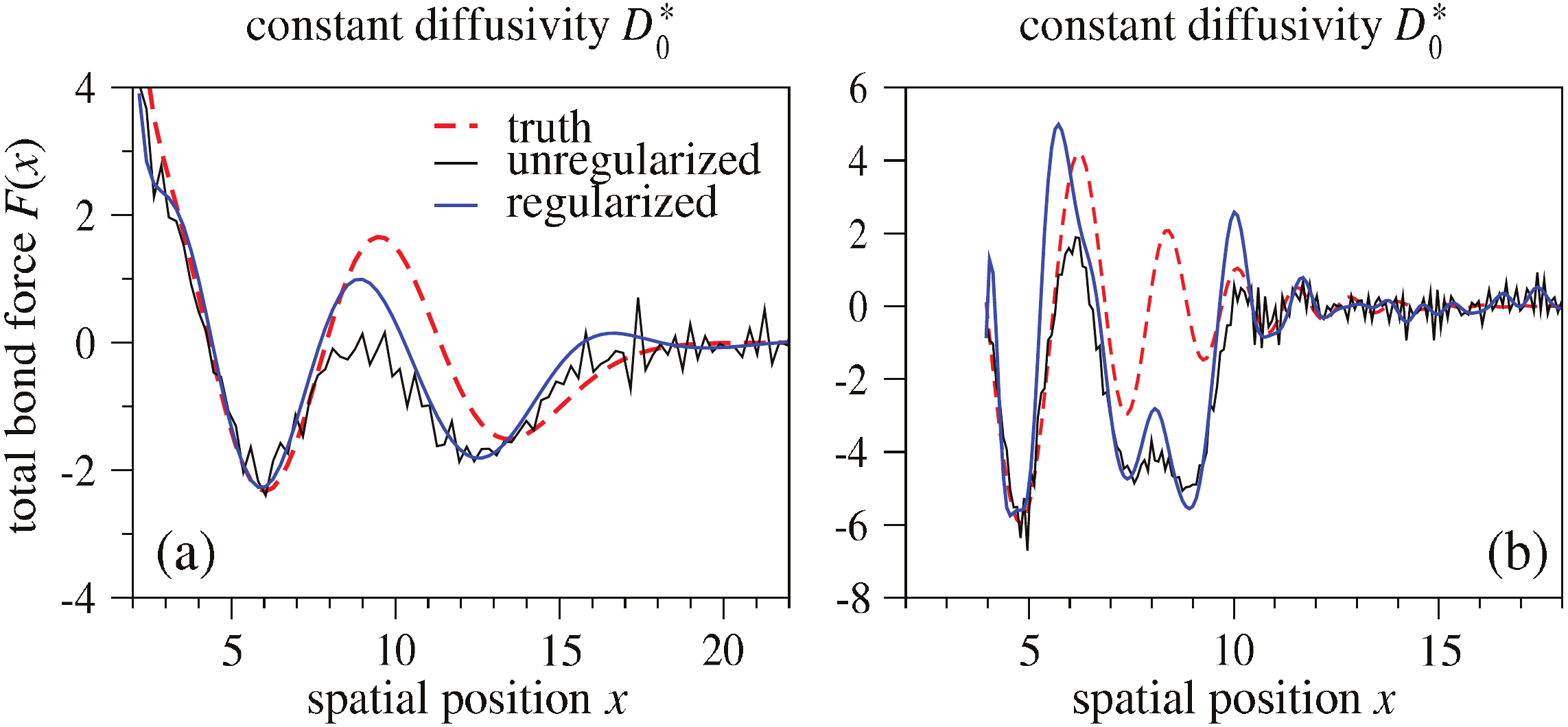}
\caption{\textbf{Failure to account for diffusivity variations.}
  Molecular bond force $F^{\star}(x) = f^{\star}(x)+F_{\textrm{ d}}(x)$
  derived from unregularized (thin black) and regularized (solid blue)
  reconstruction data simulated using a given ``ground truth'' force
  field (dashed red).  For reconstruction purposes, a constant
  diffusivity $D^{\star}_{0}$ estimated from Supplemental Eq.~\ref{eq:D0} was
  assumed.  Although regularization allows for smoother and more
  stable reconstructions, the neglect of spatial structure in $D(x)$
  leads to inaccurate results.  For example, the reconstructions in
  (a) cannot accurately determine the position of the minima, while
  those in (b) miss the minima entirely.  The errors are especially
  apparent in regions where the diffusivity is significantly different
  from the constant value (a) $D^{\star}_{0}=1.0042$ (b) $D^{\star}_{0}=0.9995$.\label{fig:fig3}}
\end{figure*}

Fig.~\ref{fig:fig3} shows reconstruction from trajectories simulated
under dynamics determined by two examples of the pair of functions
($F(x), D(x)$). These functions are explicitly given by
Supplemental Eqs.~\ref{eq:D1}--\ref{eq:f2} in the \textsc{Supplemental Methods}. The bond force shown in
Fig.~\ref{fig:fig3} corresponds to the $F(x)$ and $D(x)$ used to
generate the trajectories shown in Fig.~\ref{fig:fig2}.  Although
$D(x)$ is spatially varying, we first use a constant $D_{0}^{\star}$
obtained from Supplemental Eq.~\ref{eq:D0} in our reconstruction. Note that
regularized reconstruction (blue, dashed curves) results in smoother
and more stable recovery of $F(x)=F_{\rm d}(x)+f(x)$ compared to
unregularized recovery (thin, red curves).  However, regardless of
regularization, neglecting the true spatial dependence of $D(x)$
results in poor reconstruction of the true bond force.

%Fig.~\ref{fig:fig5} compares simultaneous reconstructions of
%the force and diffusivity across different choices of regularization.
%For comparison purposes, we include ``unregularized" reconstructions
%where the regularization matrix is diagonal, corresponding to the
%assumption that $f$ and $g$ are piecewise-constant within each of the
%$100$ bins.  Note that when the log-diffusivity $g(x)$ is
%reconstructed together, the resulting force $f^{\star}(x)$ is
%qualitatively reasonable even with out regularization.
%
%
%\begin{figure*}[t]
%\includegraphics[width=6in]{fig5}
%\caption{\textbf{Choice of regularization influences reconstruction.}
%  Simultaneous force and diffusivity reconstructions for the two sets
%  of simulated data using different regularizations. The ground truth
%  potentials and diffusivities are indicted by the dashed red curves.
%  For the first set, (a) and (b) show the unregularized piece-wise
%  constant (thin black curves) and the optimal Empirical
%  Bayes-regularized (solid blue curves) reconstructions,
%  respectively. (c) and (d) show unregularized and regularized
%  reconstructions of the diffusivity, respectively.  For the second,
%  more complex bond potential, the reconstructions are shown in
%  (e-h). Reconstructions using non-optimal regularization parameters
%  $\beta_{f,g}$, $\gamma_{f,g}$ are shown by grey dashed curves.}
%\label{fig:fig5}
%\end{figure*}

\begin{figure*}
\includegraphics[width=\linewidth]{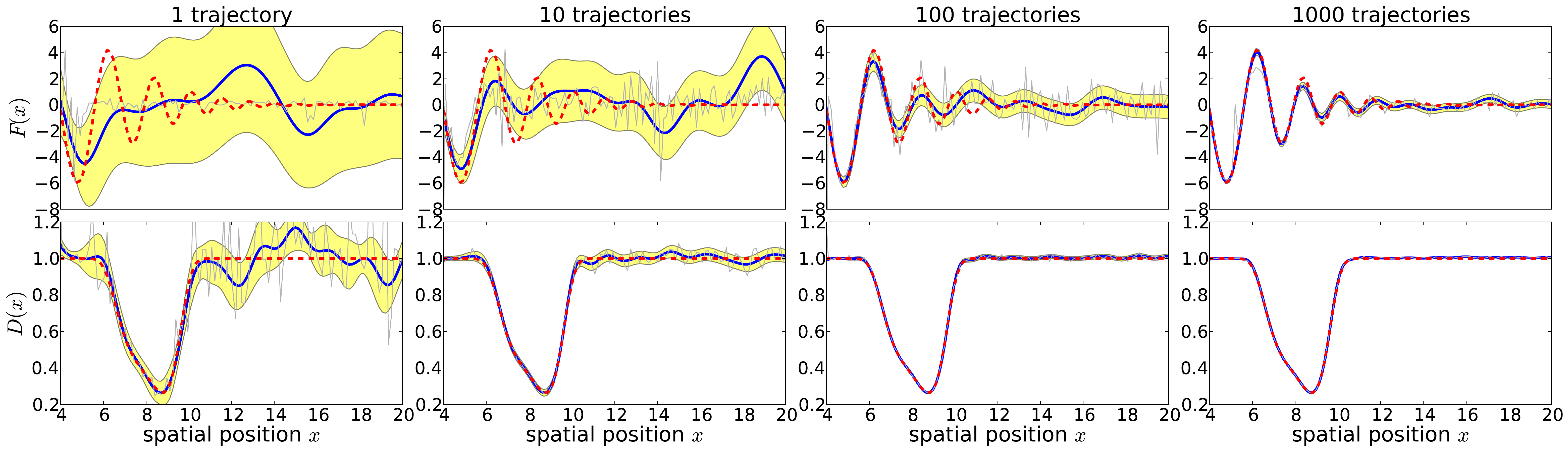}
\vspace{-1em}
\caption{\textbf{Regularized reconstruction with variable number of
    trajectories.} Reconstruction of the bond force and diffusivity
  given in Supplemental Eqs.~\ref{eq:f2}.  Shaded yellow: 95\% semiclassical
  posterior confidence interval. Grey: Unregularized bin-wise
  reconstruction. Blue: Regularized reconstructions. Optimal parameters used at 
  1000 trajectories:
  $D^*_0 =0.9995,\beta_f=19884, \gamma_f = 2.28, \beta_g=28,\gamma_g =1.02$.}
\label{fig:fig4}
\end{figure*}

Fig.~\ref{fig:fig4} demonstrates regularized reconstruction where
diffusivity variations are taken into account. It also shows how
reconstructions change as the number of observed trajectories
increases. Uncertainty quantification is also provided, where the
approximate $95\%$ posterior credible interval is shown by the
yellow-shaded region. Using physically reasonable values, we see that
a reasonable number experiments ($\sim 10^{2} - 10^{3}$) is sufficient
for simultaneous recovery of $D(x)$ and complex potentials.

%\vspace{-1.5em}
\section{Discussion and Conclusion}

We have developed a non-parametric Bayesian approach to the
simultaneous reconstruction of spatially varying bond force and
diffusivity functions directly from stochastic displacement
trajectories measured in DFS experiments.  Our approach introduces
both a path integral with explicit data terms in the energy and a
Tikhonov regularization term in the form of a prior distribution over
the functions to be recovered.  As only weak regularity conditions
based on the notion of $L^2$ integrability are used, the method is
flexible in the range of functions that can be recovered.  Moreover,
the regularization provides a formal basis for uncertainty
quantification of the reconstructed functions. The approach presented
here is versatile in that it is nonparametric, allows a broad class of
functions to be stably reconstructed, is based on the statistically
optimal principle of Bayesian inference, and can allow aggregation of
data sets from experiments performed under different conditions (such
as pulling speed $V$, device spring constant $K$, and temperature).

Our method directly uses the inherently stochastic nature of bond
trajectories to provide a likelihood formulation for use in Bayesian
inference. Hence, we are able to simultaneously and self-consistently
reconstruct two functions: the bond force and the diffusivity.  In our
example recoveries of Fig.~\ref{fig:fig3}, spatially varying
diffusivity is not included, and qualitatively incorrect reconstruction
of the bond force arises. Potentials reconstructed using constant
diffusivity can yield minima in the wrong position or miss them
altogether.  To the best of our knowledge, prior methods for
extracting information from DFS experiments, including those that
exploit work theorems
\cite{hummer2001free,hummer2005free,hummer2010free,seifert2012stochastic},
are not able to reconstruct diffusivity profiles. For this reason,
they provide an incomplete picture of the bond dynamics.

Simultaneous bond potential and diffusivity reconstruction provides
added insight into the molecular physics of the bond. Although our
test data are generated by simulations using a fixed, static ground
truth molecular potential $U(x)$ and bond force $F(x) = -\dd U(x)/\dd
x$, real molecules contain many coupled degrees of freedom. The
effective potential along the direction of bond pulling is a potential
of \emph{mean} force (PMF).  Coupling of bond displacements to other
modes of the molecule collectively contributes to a transverse
restoring force, creating a confined ``channel'' that varies in
thickness.  Such a picture of the high-dimensional potential naturally
leads to axial variations in diffusivity
\cite{zwanzig1988,best2010coordinate}. Even though our simulations
were generated from a fixed PMF $U(x)$, real data are derived from
pulling bonds that are subject to temporal fluctuations from thermal
coupling to other modes of the molecule.  Thus, both axially varying
diffusivity and thermal fluctuations are naturally subsumed in our
reconstruction of both $F(x)$ and $D(x)$ from real data.

Our approach further complements those using work theorems since
approaches using statistics of work data can be used to recover only
the mean-field solution $f^{\star}(x)$.  Moreover, our approach also
does not rely on an initial equilibrium distribution.  The
regularization operator, determined from data, incorporates the
inherent uncertainty arising from the ill-posedness of the static
inverse problem as well as the physical thermal fluctuations of the
function to be reconstructed. As the amount of data increases
(\emph{i.e.}, if more experimental trajectories are collected), the
posterior distribution for $f$ and $g$ will reflect more of the
physical uncertainty arising from the thermal fluctuations.  Our
empirically determined regularization, along with the spatially
varying ``channel'' diffusivity representation of the high-dimensional
molecular bond, provides a picture that complements the notion of a
one-dimensional PMF.

Another feature of our methodology is the inclusion of uncertainty
quantification, which provides a handle for optimizing pulling
protocols and improving recoveries.  When full trajectories are
observed and sampled, one has access to displacements in a vicinity
about any particular spatial location $x$.  The reconstruction of the
functions at $x$ utilizes trajectory measurements observed in the
neighborhood of that location, weighted by distance relative to a
characteristic length-scale $\sqrt{\gamma}$ (see
Supplemental Eq.~\ref{eq:Sigmaffyy} in the \textsc{Supplemental Methods}).  Typically,
$\sqrt{\gamma}$ spans more than one local data bin, and
self-consistent reconstructions using significantly less experimental
data are possible.  Theoretically, the recovery error of the bond
force is a function of the number of locally observed displacements,
the local diffusivity, and the net drift (Supplemental Eq.~\ref{eq:Sigmaffyy}).
In particular, the error is at a minimum when the net drift is zero,
or when the pulling force is equal and opposite to the intrinsic bond
force.

In Fig.~\ref{fig:fig4}, we empirically investigated the recovery error
as a function of the number of pulling trajectories performed.  These
plots demonstrate that features of the two functions can already be
seen with a single trajectory, are qualitatively similar to the ground
truth at 100 trajectories, and are quantitatively accurate at 1000
trajectories.  Examining Fig.~\ref{fig:fig4} in the context of
Fig.~\ref{fig:fig2}, one sees that spatial regions that are more
heavily sampled are recovered with fewer pulling experiments. By
directly observing trajectories $\xi$, one may extract information
content after a few pulls to determine optimal adjustments in $K$ and
$V$.  For example, $K$ and $V$ can be modified in order to better
probe undersampled regions of the spatial coordinate and data from
experiments using different parameters can be aggregated and used
towards the final reconstruction.

In this manuscript, we have used the regularization operator
guaranteeing infinite differentiability of the reconstructions. If
infinite differentiability is not desired, other choices are
possible~\cite{chang2014path}. We note, however, that the commonly used
Laplacian $(-\Delta)$ operator is not appropriate because its
corresponding Green's function in $\mathbb{R}^1$ does not have the
correct decay characteristics that one would expect of the bond force.
Ideally, one chooses regularization in order to represent one's prior
knowledge of the functions. For instance, one may know that the
functions should have no variations below a certain spatial scale. In
practice, this type of knowledge may not be available. 
We have utilized an empirical Bayesian approach,
thereby using the data to estimate the regularization parameters. Reconstruction
given the ``optimal" parameters within the empirical Bayesian approach
is shown by the blue curves in Fig.~\ref{fig:fig4}. Our work can be extended
to a full Bayesian treatment through use of priors on these parameters -- albeit 
at higher computational cost. Another simple extension of this work is to case of non-neglible
observation noise, by approximation of an additional path integral as in \citet{masson2014mapping,masson2009inferring}.

%XXXXXXXXXXXXXXXXXXXXXXXX

The ease of simultaneous reconstruction of $F(x)$ and $D(x)$ also
suggests that our analysis can be extended to reconstruct potential
landscapes in a few higher
dimensions~\cite{hummer2010free,suzuki2010single}, such as those
arising in catch
bonds~\cite{marshall2003direct,pereverzev2005distinctive}.  Our
approach can be readily adapted to reconstructing energy and internal
mobility profiles in extended biopolymers and multimolecular
assemblies that exhibit complex multi-minimum energy and diffusivity
profiles~\cite{koch2003dynamic,dobrovolskaia2012dynamics,thirumalai2013,titin2013}.

\section{Acknowledgments}

This material is based upon work supported by the National Science
Foundation under Agreement No. 0635561 (JC) and DMS-1021818 (TC, JC),
PHY11-25915 (KITP/UCSB), and the Army Research Office 58386MA (TC,JC).

%\section{Author Contributions}
%
%JC, PWF, and TC developed the theory and approach. PWF simulated the experiments. JC developed the numerical method. JC and TC made the figures. JC, PWF, and TC wrote and reviewed the manuscript.
%\end{acknowledgments}

\bibliographystyle{biophysj}
\bibliography{references}

\begin{thebibliography}{57}
\providecommand{\url}[1]{\texttt{#1}}
\providecommand{\urlprefix}{ }

\bibitem[Arridge(1999)]{arridge1999optical}
Arridge, S.~R., 1999.
\newblock Optical tomography in medical imaging.
\newblock \emph{Inverse Problems} 15:R41.

\bibitem[Coleman et~al.(1998)Coleman, Li, and Verma]{coleman1998reconstructing}
Coleman, T.~F., Y.~Li, and A.~Verma, 1998.
\newblock Reconstructing the unknown volatility function.
\newblock Technical report, Cornell University.

\bibitem[Ren{\`o}(2008)]{reno2008nonparametric}
Ren{\`o}, R., 2008.
\newblock Nonparametric estimation of the diffusion coefficient of stochastic
  volatility models.
\newblock \emph{Econometric Theory} 24:1174--1206.

\bibitem[Evans et~al.(1995)Evans, Ritchie, and Merkel]{evans1995sensitive}
Evans, E., K.~Ritchie, and R.~Merkel, 1995.
\newblock Sensitive force technique to probe molecular adhesion and structural
  linkages at biological interfaces.
\newblock \emph{Biophysical Journal} 68:2580--2587.

\bibitem[Heymann and Grubm{\"u}ller(2000)]{heymann2000dynamic}
Heymann, B., and H.~Grubm{\"u}ller, 2000.
\newblock Dynamic force spectroscopy of molecular adhesion bonds.
\newblock \emph{Physical Review Letters} 84:6126.

\bibitem[Merkel et~al.(1999)Merkel, Nassoy, Leung, Ritchie, and
  Evans]{merkel1999energy}
Merkel, R., P.~Nassoy, A.~Leung, K.~Ritchie, and E.~Evans, 1999.
\newblock Energy landscapes of receptor--ligand bonds explored with dynamic
  force spectroscopy.
\newblock \emph{Nature} 397:50--53.

\bibitem[Neuman and Nagy(2008)]{neuman2008single}
Neuman, K.~C., and A.~Nagy, 2008.
\newblock Single-molecule force spectroscopy: optical tweezers, magnetic
  tweezers and atomic force microscopy.
\newblock \emph{Nature methods} 5:491.

\bibitem[Lang et~al.(2004)Lang, Fordyce, Engh, Neuman, and
  Block]{lang2004simultaneous}
Lang, M.~J., P.~M. Fordyce, A.~M. Engh, K.~C. Neuman, and S.~M. Block, 2004.
\newblock Simultaneous, coincident optical trapping and single-molecule
  fluorescence.
\newblock \emph{Nature Methods} 1:133--139.

\bibitem[Hinterdorfer and Dufr{\^e}ne(2006)]{hinterdorfer2006detection}
Hinterdorfer, P., and Y.~F. Dufr{\^e}ne, 2006.
\newblock Detection and localization of single molecular recognition events
  using atomic force microscopy.
\newblock \emph{Nature methods} 3:347--355.

\bibitem[Rawicz et~al.(2008)Rawicz, Smith, McIntosh, Simon, and
  Evans]{rawicz2008elasticity}
Rawicz, W., B.~Smith, T.~McIntosh, S.~Simon, and E.~Evans, 2008.
\newblock Elasticity, strength, and water permeability of bilayers that contain
  raft microdomain-forming lipids.
\newblock \emph{Biophysical journal} 94:4725--4736.

\bibitem[Koch and Wang(2003)]{koch2003dynamic}
Koch, S.~J., and M.~D. Wang, 2003.
\newblock Dynamic force spectroscopy of protein-{DNA} interactions by unzipping
  {DNA}.
\newblock \emph{Physical Review Letters} 91:028103.

\bibitem[Jobst et~al.(2013)Jobst, Schoeler, Malinowska, and
  Nash]{jobst2013investigating}
Jobst, M.~A., C.~Schoeler, K.~Malinowska, and M.~A. Nash, 2013.
\newblock Investigating Receptor-ligand Systems of the Cellulosome with
  AFM-based Single-molecule Force Spectroscopy.
\newblock \emph{JoVE (Journal of Visualized Experiments)} e50950--e50950.

\bibitem[Maitra and Arya(2010)]{maitra2010model}
Maitra, A., and G.~Arya, 2010.
\newblock Model accounting for the effects of pulling-device stiffness in the
  analyses of single-molecule force measurements.
\newblock \emph{Physical Review Letters} 104:108301.

\bibitem[Hummer and Szabo(2001)]{hummer2001free}
Hummer, G., and A.~Szabo, 2001.
\newblock Free energy reconstruction from nonequilibrium single-molecule
  pulling experiments.
\newblock \emph{Proceedings of the National Academy of Sciences} 98:3658--3661.

\bibitem[Hummer and Szabo(2010)]{hummer2010free}
Hummer, G., and A.~Szabo, 2010.
\newblock Free energy profiles from single-molecule pulling experiments.
\newblock \emph{Proceedings of the National Academy of Sciences}
  107:21441--21446.

\bibitem[Rief et~al.(1997)Rief, Oesterhelt, Heymann, and Gaub]{rief1997single}
Rief, M., F.~Oesterhelt, B.~Heymann, and H.~E. Gaub, 1997.
\newblock Single molecule force spectroscopy on polysaccharides by atomic force
  microscopy.
\newblock \emph{Science} 275:1295--1297.

\bibitem[Puchner and Gaub(2009)]{puchner2009force}
Puchner, E.~M., and H.~E. Gaub, 2009.
\newblock Force and function: probing proteins with AFM-based force
  spectroscopy.
\newblock \emph{Current opinion in structural biology} 19:605--614.

\bibitem[Fernandez and Li(2004)]{fernandez2004force}
Fernandez, J.~M., and H.~Li, 2004.
\newblock Force-clamp spectroscopy monitors the folding trajectory of a single
  protein.
\newblock \emph{Science} 303:1674--1678.

\bibitem[Dobrovolskaia and Arya(2012)]{dobrovolskaia2012dynamics}
Dobrovolskaia, I.~V., and G.~Arya, 2012.
\newblock Dynamics of forced nucleosome unraveling and role of nonuniform
  histone-{DNA} interactions.
\newblock \emph{Biophysical Journal} 103:989--998.

\bibitem[Ros et~al.(2004)Ros, Eckel, Bartels, Sischka, Baumgarth, Wilking,
  P{\"u}hler, Sewald, Becker, and Anselmetti]{ros2004single}
Ros, R., R.~Eckel, F.~Bartels, A.~Sischka, B.~Baumgarth, S.~D. Wilking,
  A.~P{\"u}hler, N.~Sewald, A.~Becker, and D.~Anselmetti, 2004.
\newblock Single molecule force spectroscopy on ligand--{DNA} complexes: from
  molecular binding mechanisms to biosensor applications.
\newblock \emph{Journal of biotechnology} 112:5--12.

\bibitem[Rief et~al.(1999)Rief, Pascual, Saraste, and Gaub]{rief1999single}
Rief, M., J.~Pascual, M.~Saraste, and H.~E. Gaub, 1999.
\newblock Single molecule force spectroscopy of spectrin repeats: low unfolding
  forces in helix bundles.
\newblock \emph{Journal of molecular biology} 286:553--561.

\bibitem[Clausen-Schaumann et~al.(2000)Clausen-Schaumann, Seitz, Krautbauer,
  and Gaub]{clausen2000force}
Clausen-Schaumann, H., M.~Seitz, R.~Krautbauer, and H.~E. Gaub, 2000.
\newblock Force spectroscopy with single bio-molecules.
\newblock \emph{Current Opinion in Chemical Biology} 4:524--530.

\bibitem[Helenius et~al.(2008)Helenius, Heisenberg, Gaub, and
  Muller]{helenius2008single}
Helenius, J., C.-P. Heisenberg, H.~E. Gaub, and D.~J. Muller, 2008.
\newblock Single-cell force spectroscopy.
\newblock \emph{Journal of Cell Science} 121:1785--1791.

\bibitem[Anselmetti et~al.(2007)Anselmetti, Hansmeier, Kalinowski, Martini,
  Merkle, Palmisano, Ros, Schmied, Sischka, and
  T{\"o}nsing]{anselmetti2007analysis}
Anselmetti, D., N.~Hansmeier, J.~Kalinowski, J.~Martini, T.~Merkle,
  R.~Palmisano, R.~Ros, K.~Schmied, A.~Sischka, and K.~T{\"o}nsing, 2007.
\newblock Analysis of subcellular surface structure, function and dynamics.
\newblock \emph{Analytical and Bioanalytical Chemistry} 387:83--89.

\bibitem[Benoit et~al.(2000)Benoit, Gabriel, Gerisch, and
  Gaub]{benoit2000discrete}
Benoit, M., D.~Gabriel, G.~Gerisch, and H.~E. Gaub, 2000.
\newblock Discrete interactions in cell adhesion measured by single-molecule
  force spectroscopy.
\newblock \emph{Nature Cell Biology} 2:313--317.

\bibitem[Evans and Calderwood(2007)]{evans2007forces}
Evans, E.~A., and D.~A. Calderwood, 2007.
\newblock Forces and bond dynamics in cell adhesion.
\newblock \emph{Science} 316:1148--1153.

\bibitem[Dudko et~al.(2008)Dudko, Hummer, and Szabo]{dudko2008theory}
Dudko, O.~K., G.~Hummer, and A.~Szabo, 2008.
\newblock Theory, analysis, and interpretation of single-molecule force
  spectroscopy experiments.
\newblock \emph{Proceedings of the National Academy of Sciences}
  105:15755--15760.

\bibitem[Dudko(2009)]{dudko2009single}
Dudko, O.~K., 2009.
\newblock Single-molecule mechanics: New insights from the
  escape-over-a-barrier problem.
\newblock \emph{Proceedings of the National Academy of Sciences}
  106:8795--8796.

\bibitem[Freund(2009)]{freund2009characterizing}
Freund, L., 2009.
\newblock Characterizing the resistance generated by a molecular bond as it is
  forcibly separated.
\newblock \emph{Proceedings of the National Academy of Sciences}
  106:8818--8823.

\bibitem[Fuhrmann et~al.(2008)Fuhrmann, Anselmetti, Ros, Getfert, and
  Reimann]{fuhrmann2008refined}
Fuhrmann, A., D.~Anselmetti, R.~Ros, S.~Getfert, and P.~Reimann, 2008.
\newblock Refined procedure of evaluating experimental single-molecule force
  spectroscopy data.
\newblock \emph{Physical Review E} 77:031912.

\bibitem[Evstigneev and Reimann(2003)]{evstigneev2003dynamic}
Evstigneev, M., and P.~Reimann, 2003.
\newblock Dynamic force spectroscopy: optimized data analysis.
\newblock \emph{Physical Review E} 68:045103.

\bibitem[Shapiro and Qian(1997)]{shapiro1997quantitative}
Shapiro, B.~E., and H.~Qian, 1997.
\newblock A quantitative analysis of single protein-ligand complex separation
  with the atomic force microscope.
\newblock \emph{Biophysical chemistry} 67:211--219.

\bibitem[Hummer and Szabo(2005)]{hummer2005free}
Hummer, G., and A.~Szabo, 2005.
\newblock Free energy surfaces from single-molecule force spectroscopy.
\newblock \emph{Accounts of Chemical Research} 38:504--513.

\bibitem[Balsera et~al.(1997)Balsera, Stepaniants, Izrailev, Oono, and
  Schulten]{balsera1997reconstructing}
Balsera, M., S.~Stepaniants, S.~Izrailev, Y.~Oono, and K.~Schulten, 1997.
\newblock Reconstructing potential energy functions from simulated
  force-induced unbinding processes.
\newblock \emph{Biophysical Journal} 73:1281--1287.

\bibitem[Woodside and Block(2014)]{block_review}
Woodside, M.~T., and S.~M. Block, 2014.
\newblock Reconstructing Folding Energy Landscapes by Single-Molecule Force
  Spectroscopy.
\newblock \emph{Annual Review of Biophysics} 43:19--39.

\bibitem[T{\"u}rkcan et~al.(2012)T{\"u}rkcan, Alexandrou, and
  Masson]{turkcan2012bayesian}
T{\"u}rkcan, S., A.~Alexandrou, and J.-B. Masson, 2012.
\newblock A Bayesian inference scheme to extract diffusivity and potential
  fields from confined single-molecule trajectories.
\newblock \emph{Biophysical Journal} 102:2288--2298.

\bibitem[Masson et~al.(2014)Masson, Dionne, Salvatico, Renner, Specht, Triller,
  and Dahan]{masson2014mapping}
Masson, J.-B., P.~Dionne, C.~Salvatico, M.~Renner, C.~G. Specht, A.~Triller,
  and M.~Dahan, 2014.
\newblock Mapping the energy and diffusion landscapes of membrane proteins at
  the cell surface using high-density single-molecule imaging and Bayesian
  Inference: application to the multiscale dynamics of glycine receptors in the
  neuronal membrane.
\newblock \emph{Biophysical Journal} 106:74--83.

\bibitem[Schuss(2011)]{schuss2011nonlinear}
Schuss, Z., 2011.
\newblock Nonlinear filtering and optimal phase tracking, volume 180.
\newblock Springer.

\bibitem[Alemany et~al.(2012)Alemany, Mossa, Junier, and
  Ritort]{alemany2012experimental}
Alemany, A., A.~Mossa, I.~Junier, and F.~Ritort, 2012.
\newblock Experimental free-energy measurements of kinetic molecular states
  using fluctuation theorems.
\newblock \emph{Nature Physics} 8:688--694.

\bibitem[Seifert(2012)]{seifert2012stochastic}
Seifert, U., 2012.
\newblock Stochastic thermodynamics, fluctuation theorems and molecular
  machines.
\newblock \emph{Reports on Progress in Physics} 75:126001.

\bibitem[Zwanzig(1988)]{zwanzig1988}
Zwanzig, R., 1988.
\newblock Diffusion in a rough potential.
\newblock \emph{Proceedings of the National Academy of Sciences} 85:2029--2030.

\bibitem[Best and Hummer(2010)]{best2010coordinate}
Best, R.~B., and G.~Hummer, 2010.
\newblock Coordinate-dependent diffusion in protein folding.
\newblock \emph{Proceedings of the National Academy of Sciences}
  107:1088--1093.

\bibitem[Fok and Chou(2010)]{fok2010reconstruction}
Fok, P.-W., and T.~Chou, 2010.
\newblock Reconstruction of potential energy profiles from multiple rupture
  time distributions.
\newblock \emph{Proceedings of the Royal Society A: Mathematical, Physical and
  Engineering Science} 466:3479--3499.

\bibitem[Schuss(2009)]{schuss2009theory}
Schuss, Z., 2009.
\newblock Theory and applications of stochastic processes: an analytical
  approach, volume 170.
\newblock Springer.

\bibitem[Lemm et~al.(2000)Lemm, Uhlig, and Weiguny]{lemm2000bayesian}
Lemm, J., J.~Uhlig, and A.~Weiguny, 2000.
\newblock Bayesian approach to inverse quantum statistics.
\newblock \emph{Physical review letters} 84:2068.

\bibitem[Chang et~al.(2014)Chang, Savage, and Chou]{chang2014path}
Chang, J.~C., V.~M. Savage, and T.~Chou, 2014.
\newblock A Path-Integral Approach to {{B}}ayesian Inference for Inverse
  Problems Using the Semiclassical Approximation.
\newblock \emph{Journal of Statistical Physics} 157:582--602.

\bibitem[En{\ss}lin et~al.(2009)En{\ss}lin, Frommert, and
  Kitaura]{ensslin2009information}
En{\ss}lin, T.~A., M.~Frommert, and F.~S. Kitaura, 2009.
\newblock Information field theory for cosmological perturbation reconstruction
  and nonlinear signal analysis.
\newblock \emph{Physical Review D} 80:105005.

\bibitem[Cotter et~al.(2009)Cotter, Dashti, Robinson, and
  Stuart]{cotter2009bayesian}
Cotter, S., M.~Dashti, J.~Robinson, and A.~Stuart, 2009.
\newblock {Bayesian inverse problems for functions and applications to fluid
  mechanics}.
\newblock \emph{Inverse Problems} 25:115008.

\bibitem[Heuett et~al.(2012)Heuett, Miller~III, Racette, Holloszy, Chow, and
  Periwal]{heuett2012bayesian}
Heuett, W.~J., B.~V. Miller~III, S.~B. Racette, J.~O. Holloszy, C.~C. Chow, and
  V.~Periwal, 2012.
\newblock Bayesian Functional Integral Method for Inferring Continuous Data
  from Discrete Measurements.
\newblock \emph{Biophysical Journal} 102:399--406.

\bibitem[Farmer(2007)]{farmer2007bayesian}
Farmer, C., 2007.
\newblock {Bayesian field theory applied to scattered data interpolation and
  inverse problems}.
\newblock \emph{Algorithms for Approximation} 147--166.

\bibitem[Stuart(2010)]{stuart2010inverse}
Stuart, A., 2010.
\newblock {Inverse problems: a Bayesian perspective}.
\newblock \emph{Acta Numerica} 19:451--559.

\bibitem[Masson et~al.(2009)Masson, Casanova, T{\"u}rkcan, Voisinne, Popoff,
  Vergassola, and Alexandrou]{masson2009inferring}
Masson, J.-B., D.~Casanova, S.~T{\"u}rkcan, G.~Voisinne, M.-R. Popoff,
  M.~Vergassola, and A.~Alexandrou, 2009.
\newblock Inferring maps of forces inside cell membrane microdomains.
\newblock \emph{Physical review letters} 102:048103.

\bibitem[Suzuki and Dudko(2010)]{suzuki2010single}
Suzuki, Y., and O.~K. Dudko, 2010.
\newblock Single-molecule rupture dynamics on multidimensional landscapes.
\newblock \emph{Physical Review Letters} 104:048101.

\bibitem[Marshall et~al.(2003)Marshall, Long, Piper, Yago, McEver, and
  Zhu]{marshall2003direct}
Marshall, B.~T., M.~Long, J.~W. Piper, T.~Yago, R.~P. McEver, and C.~Zhu, 2003.
\newblock Direct observation of catch bonds involving cell-adhesion molecules.
\newblock \emph{Nature} 423:190--193.

\bibitem[Pereverzev et~al.(2005)Pereverzev, Prezhdo, Thomas, and
  Sokurenko]{pereverzev2005distinctive}
Pereverzev, Y.~V., O.~V. Prezhdo, W.~E. Thomas, and E.~V. Sokurenko, 2005.
\newblock Distinctive features of the biological catch bond in the jump-ramp
  force regime predicted by the two-pathway model.
\newblock \emph{Physical Review E} 72:010903.

\bibitem[Hinczewski et~al.(2013)Hinczewski, Gebhardt, Reif, and
  Thirumalai]{thirumalai2013}
Hinczewski, M., J.~C.~M. Gebhardt, M.~Reif, and D.~Thirumalai, 2013.
\newblock From mechanical folding trajectories to intrinsic energy landscapes
  of biopolymers.
\newblock \emph{Proceedings of the National Academy of Sciences}
  110:4500--4505.

\bibitem[Rico et~al.(2013)Rico, Gonzalez, Casuso, Puig-Vidal, and
  Scheuring]{titin2013}
Rico, F., L.~Gonzalez, I.~Casuso, M.~Puig-Vidal, and S.~Scheuring, 2013.
\newblock High-Speed Force Spectroscopy Unfolds Titin at the Velocity of
  Molecular Dynamics Simulations.
\newblock \emph{Science} 342:741--743.

\end{thebibliography}
\newpage
\clearpage
\onecolumngrid
\renewcommand\appendixtocname{Supplemental Methods}
\renewcommand\appendixpagename{Supplemental Methods}
%\appendix
%\appendixpage
\textbf{SUPPLEMENTAL METHODS}
\setcounter{section}{0}
\setcounter{page}{1}

\setcounter{equation}{0}
\renewcommand{\theequation}{S\arabic{equation}}
\setcounter{figure}{0}
\renewcommand{\thefigure}{S\arabic{figure}}
\renewcommand{\thesection}{Supplemental Methods\ \arabic{section}}

\section{Functions used in our examples}

\begin{figure*}[b]
\centering
\includegraphics[width=\linewidth]{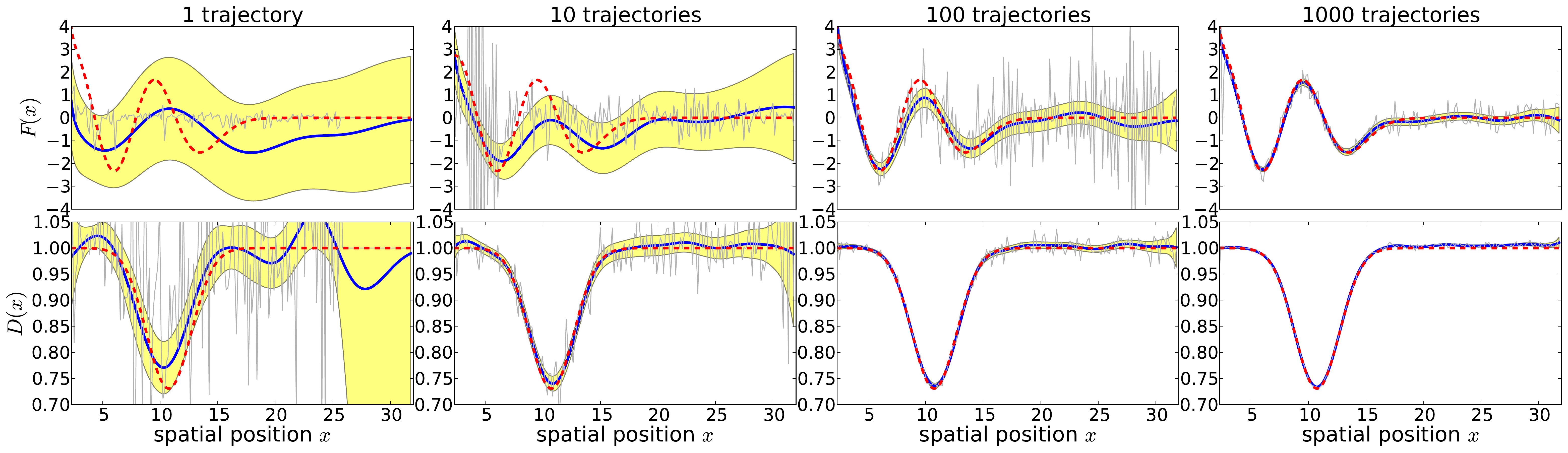}
\caption{\textbf{Regularized reconstruction with variable number of
    trajectories.} Reconstruction of the bond force and diffusivity
  given in Eqs.~\ref{eq:D1},~\ref{eq:f1}.  Shaded yellow: 95\%
  semiclassical posterior confidence interval. Grey: Unregularized
  bin-wise reconstruction.}
\label{fig:Sfig1}
\end{figure*}

In our examples we used two feature-rich pairs of diffusivity $D$ and bond force
$F=F_d + f$.  In all cases $F_{\rm d}(x) =
\left(\frac{x}{2}\right)^{-6}$.  The trajectories shown in 
Fig.~\ref{fig:fig2} were generated using

\begin{equation}
D(x) = 1-\frac{x^2}{400}\exp\left(-\frac{(x-10)^2}{8}\right), \label{eq:D1}
\end{equation} 

\begin{align}
f(x) &=
\frac{3\sqrt{x}}{10}\exp\left(-\frac{(x-10)^2}{12}\right) -
\frac{x^2(5-x)}{35}\exp\left(-\frac{(x-5)^2}{14}\right) +
\frac{x^2(2-x)}{10}\exp\left(-\frac{(x-2)^2}{16}\right) \nonumber\\
&\qquad+ \frac{8x}{5}
\exp\left(-\frac{(x-2)^2}{16}\right) -
\frac{2x}{5}\exp\left(-\frac{(x-5)^2}{14}\right) +
\frac{x^{3/2}(10-x)}{30}\exp\left(-\frac{(x-10)^2}{12}\right).\label{eq:f1}
\end{align}
These forms were also used in the reconstruction of $F(x)$
shown in Fig.~\ref{fig:fig3}(a). Fig.~\ref{fig:Sfig1}  shows
simultaneous reconstructions of $D(x)$ and $F(x)$ defined in
Eqs.~\ref{eq:D1} and \ref{eq:f1}.

In Figs.~\ref{fig:fig3}(b) and \ref{fig:fig4}, we considered a different 
diffusivity profile and a more 
complex potential:
\begin{equation}
D(x) = 1-\frac{x^2}{100}\exp\left(-\frac{(x-10)^4}{8}\right), \quad 
f(x) = 10\sin(x^2/5)\exp(-x^2/45).\label{eq:f2}
\end{equation}

\section{Transition probabilities}\label{sec:transitions}
Assuming It\^{o} calculus, the Brownian motion is described through the SDE
\begin{equation}
\dd X = A( X,t)\dd t + \sqrt{2D( X)}\dd W \label{eq:sde1}
\end{equation}
where $W$ is the Wiener white noise process, $D(x)$ is the diffusivity, and
\begin{equation}
A(x,t) = D(x)\partial_x(-\Phi(x,t) + \log D(x))
\end{equation}
In order to compute this quantity we first consider the short-time
solution of the SDE using It\^{o} rules. Let $h\to0$ be a small
timestep. Then, we have
\begin{align}
\int_{t=t_0}^{t=t_0+h} \dd x &=  X(t_0+h)- X(t_0) \nonumber\\
&=\int_{t=t_0}^{t=t_0+h} A( X(t),t)\dd t + \int_{t=t_0}^{t=t_0+h} \sqrt{2D( X)}\dd W \nonumber\\
&= A( X(t_0),t_0)h+ z\sqrt{2D( X(t_0))h}  +\mathcal{O}(h^{3/2}), 
\end{align}
where $z$ is a standard normal random variable. In the limit as $h\to0$, we can write
\begin{equation}
X(t_0+h)\sim \mathcal{N}\Big(X(t_0)+A(X(t_0),t_0)h, 2D(X(t_0))h \Big), 
\end{equation}
which implies that as $h\to 0$,
\begin{equation}
\Pr\Big( X(t_0+h)\Big|  X(t_0) \Big) = \left[\frac{1}{4\pi D( X(t_0))h} \right]^{1/2}\exp\left\{ -\frac{\Big[ X(t_0+h)-x(t_0)-A\big( X(t_0),t_0\big)h\Big]^2}{4D( X(t_0))h} \right\}.
\end{equation}
In practice, observations of the trajectory positions are taken with noise. Assuming that the noise is i.i.d. Gaussian with zero mean and variance $\sigma^2$,
the likelihood of observing a particular trajectory $\boldsymbol{\xi}=\xi_0,\xi_1,\ldots$ sampled at time increments of width $\delta t$ given a particular choice of $f,g$ is the product of the probabilities of observing each of the transitions, or
\begin{align}
\pi(\boldsymbol{\xi}|f,g) &= \int \Pr(\xi_0,X_0)dX_0 \prod_{j=0} \Pr(\xi_{j+1},X_{j+1} |f,g,X_j,\sigma^2) \dd{X_{j+1}} \nonumber\\
&\approx \exp\left\{-\frac{1}{2}\sum_j\left[ \frac{(\xi_{j+1}-\xi_j-A(\xi_j,t_j)\delta t)^2}{2 D(\xi_j)\delta t } +\log(4\pi D(\xi_j)\delta t) \right] \right\},
\end{align} 
where the integrals with respect to $X_j$ have been evaluated using Laplace's approximation under the assumption that
\begin{equation}
\frac{\sigma^2}{2D(x)\delta t} \ll 1, \qquad\forall x.
\end{equation}
In the case where there are multiple independent trajectories $\Xi = \{ \boldsymbol{\xi}^{(\alpha)}\}$,
\begin{equation}
\pi({\Xi}|f,g) =  \exp\left\{-\frac{1}{2}\sum_{j,\alpha}\left[ \frac{(\xi^{(\alpha)}_{j+1}-\xi^{(\alpha)}_j-A(\xi^{(\alpha)}_j,t_j)\delta t)^2}{2 D(\xi^{(\alpha)}_j)\delta t} +\log(4\pi D(\xi^{(\alpha)}_j)\delta t) \right] \right\}.\label{eq:likelihood}
\end{equation}

\section{Empirical estimation of the background diffusivity and cantilever stiffness constant} 
If the background diffusivity $D_0$ is unknown, it can be estimated
directly from observations of the Brownian motion. Similarly, the cantilever spring constant $K$, 
usually determined by one of several procedures, can be refined.

The observed displacements in the trajectories originating at position $x$
at time $t$ are normally distributed with mean $A(x,t)\delta
t$ and variance $2D(x)\delta t$. In the large $x$ limit, $A(x,t)\to
D_0 K(L(t)-x)$, and $D(x)\to D_0$. One may then simply estimate the background
diffusivity $D_0$ and spring constant $K$ using the displacements from all 
trajectories that extend past a critical cutoff separation $x_c$.  The negative log-likelihood function
for these observations is
\begin{align}
\mathcal{L}\equiv-\log\pi(\{\xi_{\geq x_c}\}) = \frac{1}{2}\sum_{\obsj\geq x_c}\left\{ \log\left(4\pi D_0\delta t\right) 
+\frac{ [\obs{j+1}{\alpha}-\obsj - D_0 K(L(t_j)-\obsj )\delta t]^2}{2D_0\delta t}
\right\}.\label{eq:KD0like}
\end{align}
%Additionally, we impose a prior distribution on the spring constant $K$ about its 
%measured value $K_0$,
%\begin{equation}
%\pi(K\vert K_0) = \frac{1}{\sqrt{2\pi(K_0/10)^2}}\exp\left[-\frac{1}{2(K_0/10)^2}\left(K-K_0 \right)^2 \right],
%\end{equation}
%to reflect the fact that it has been measured to an  error of $10\%$. 
The optimal parameters
$D_0$ and $K$ can be found through maximization of Eq.~\ref{eq:KD0like}.
 This procedure is accomplished by solving the system of equations
\begin{align}
\frac{\partial\mathcal{L}}{\partial K }&=K\left[ \frac{D_0\delta t}{2}\sum_{\mathclap{\xi_j^{(\alpha)}\geq x_c}}(L(t_j)-\obsj)^2\right] -\sum_{\mathclap{\xi_j^{(\alpha)}\geq x_c}}\frac{(\obs{j+1}{\alpha}-\obsj)(L(t_j)-\obsj)}{2} = 0
\end{align}  

\begin{align}
\frac{\partial\mathcal{L}}{\partial D_0 }&= \sum_{\mathclap{\xi_j^{(\alpha)}\geq x_c}}
\left\{\frac{1-K(L(t_j)-\obsj )(\obs{j+1}{\alpha}-\obsj-D_0K(L(t_j)-\obsj )\delta t )}{2D_0}\right\}\nonumber\\
&- \sum_{\mathclap{\xi_j^{(\alpha)}\geq x_c}}\left\{\frac{(\obs{j+1}{\alpha}-\obsj-D_0K(L(t_j)-\obsj)\delta t)^2}{4D_0^2\delta t} \right\} = 0.
\end{align}

The maximum likelihood estimates for $D_0$ and $K$ are
\begin{align}
K^* = \left[\ \ \sum_{\mathclap{\xi_j^{(\alpha)}\geq x_c}}\frac{(\obs{j+1}{\alpha}-\obsj)(L(t_j)-\obsj)}{2}\right] \Bigg/ \left[ \frac{D^*_0\delta t}{2}\sum_{\mathclap{\xi_j^{(\alpha)}\geq x_c}}(L(t_j)-\obsj)^2\right]\label{eq:Kstar},
\end{align}

 {\small
\begin{align}%\sum_{\mathclap{\xi_j^{(\alpha)}\geq x_c}}
D_0^{*}& =\left\{\sqrt{  {K*}^2\sum_{\mathclap{\xi_j^{(\alpha)}\geq x_c}}(d_j^{(\alpha)})^2 \sum_{\mathclap{\xi_j^{(\alpha)}\geq x_c}}(\eta_j^{(\alpha)})^2 +(\sum_{\mathclap{\xi_j^{(\alpha)}\geq x_c}}1)^2    }\quad-\sum_{\mathclap{\xi_j^{(\alpha)}\geq x_c}}1 \right\}\Bigg/\Bigg[\ \delta t {K^{*2}\sum_{\mathclap{\xi_j^{(\alpha)}\geq x_c}} (d_j^{(\alpha)})^2 } \Bigg],
\label{eq:D0}
\end{align}
}
where $d_j^{(\alpha)} = (L(t_j)-\xi_j^{(\alpha)})$ and 
$\eta_j^{(\alpha)} = \xi_{j+1}^{(\alpha)}-\xi_j^{(\alpha)}$. These equations can be solved by Newton-Raphson iteration.

\section{Inference}\label{sec:eulerlagrange}

\subsection{Euler-Lagrange equations}
The Euler-Lagrange equations for the Information Hamiltonian
%(Eq.~\ref{eq:hamiltonian}) 
are obtained by computing variational
derivatives with respect to the functions $f(y), g(y)$ and setting
them to zero. Using the Dirac delta function 
we rewrite the Information Hamiltonian in the integral form
{\small {\begin{align}
H\left[f,g\mid\Xi\right]&=\frac{1}{2}\int_0^\infty f(y)R_f(-\Delta)f(y)\d{y} +\frac{1}{2}\int_0^\infty g(y)R_g(-\Delta)g(y)\d{y} \nonumber\\
&\quad+\frac{1}{2}\sum_{\alpha,j}\int \delta(y-\xi^{(\alpha)}_j)  \log D(y)\d{y}+ \sum_{\alpha,j} \int  \delta(y-\xi^{(\alpha)}_j) \frac{{\left(\xi^{(\alpha)}_{j+1}-\xi^{(\alpha)}_j-A(y,t_j)\delta t\right)^2} }{4D(y)\delta t}\d{y}. 
\label{eq:hamiltonian1}
\end{align}}}
To calculate the variational derivatives, the following relations
\begin{equation}
\partial_g D =D \qquad \partial_g A = A\qquad\partial_{g^\prime} A = D \qquad \partial_f A = D
\end{equation}
will prove useful.
Using these identities, we straightforwardly take variations of $H$
with respect to both $f(y)$ and $g(y)$ to find

{\small
\begin{equation}
\frac{\delta H}{\delta f(y)}=R_f(-\Delta)f(y) - \frac{1}{2}\sum_{\alpha,j} \delta(y-\xi_j^\alpha) \left[ \xi_{j+1}^{(\alpha)}-\xi_j^{(\alpha)}-A(y,t)\delta t \right] \label{eq:deltaf}
\end{equation}
\begin{align}
\frac{\delta H}{\delta g(y)}&= R_g(-\Delta)g(y)+\frac{1}{2}\sum_{\alpha,j}\frac{\partial}{\partial y}\left[ \delta(y-\xi_{j}^{(\alpha)} ) (\xi_{j+1}^{(\alpha)} -\xi_j^{(\alpha)} -A(y,t_j)\delta t)  \right]  \nonumber \\
& +\frac{1}{2}\sum_{\alpha,j} \delta(y-\xi_j^{(\alpha)})\Bigg\{1-\frac{ (\xi_{j+1}^{(\alpha)} -\xi_j^{(\alpha)})^2 - A^2(y,t_j) (\delta t)^2 }{2D(y) \delta t} \Bigg\}.
\label{eq:dD}
\end{align}
}

Equations \ref{eq:deltaf} and \ref{eq:dD}, set to zero, yield the
Euler-Lagrange equations. We solve these equations using their
corresponding Greens functions. The operators $R_{f,g}$ have the associated free-space Green's function $G_{\infty}(x,y)=
\beta \exp\left[-(x-y)^2/(2\gamma)\right]$.
The parameter $\beta>0$ acts like an inverse temperature and controls
the magnitude of the variability found in a field. The parameter
$\gamma>0$ is a spatial scale parameter, strongly penalizing
variations at length scales at or smaller than
$\mathcal{O}(\sqrt\gamma)$. Since recovery is over the positive part
of the real line, and we are fixing the function values for $f$ and
$g$ to zero at $x=0$, we enforce the condition that variations in the
functions $f$ and $g$ are not correlated to $f(0)$ and $g(0)$,
respectively. Hence, we use the method of images to enforce an
absorbing boundary condition at $x=0$ and for $f$ and $g$ write the
full Green's function as

\begin{equation}
G(x,y) = \beta\exp\left[-\frac{(x-y)^2}{2\gamma}\right]-
\beta\exp\left[-\frac{(x+y)^2}{2\gamma}\right].
\label{GREENS}
\end{equation}
The Green's function for the regularization operator defines the {\it
a priori} spatial variation in the functions that make up the space of
functions described by the distributions $\pi(f), \pi(g)$. 

 The solution to the Euler-Lagrange equations can be 
formally written as a linear equation 
for $f(y)$ {\small
\begin{align}
0=f(y)-\frac{1}{2}\sum_{\alpha,j} G_f(y,\xi_{j}^{(\alpha)})\Bigg[ \frac{\xi_{j+1}^{(\alpha)}-\xi_{j}^{(\alpha)}}{\delta t} -D(\xi_{j}^{(\alpha)})\left( f(\xi_{j}^{(\alpha)})+m(\xi_{j}^{(\alpha)},t_j) \right) -D^\prime(\xi_{j}^{(\alpha)}) \Bigg]\delta t, \label{eq:deltafsliced}
\end{align}}
and a nonlinear equation for $g(y)$

{\small
\begin{align}
&\lefteqn{0=g(y)-\frac{1}{2}\sum_{\alpha,j} \left[\frac{\partial}{\partial z}G_g(y,\xi_j^{(\alpha)})\right] \Bigg[\xi^{(\alpha)}_{j+1}-\xi^{(\alpha)}_{j}-A(\xi_j^{(\alpha)},t_j)\delta t\Bigg] }\nonumber\\
& +\frac{1}{2}\sum_{\alpha,j}  G_g(y,\xi_j^{(\alpha)})\Bigg\{1-\frac{ (\xi_{j+1}^{(\alpha)} -\xi_j^{(\alpha)})^2 - A^2(\xi_j^{(\alpha)},t_j) (\delta t)^2 }{2D(\xi_j^{(\alpha)}) \delta t} \Bigg\}\label{eq:dDsliced}
\end{align}}
where $G_f$ is the Green's function for $R_f(-\Delta)$ and $G_g$ is
the Green's function for $R_g(-\Delta)$, and
\begin{equation}
m(y,t) \equiv F_d(y)+K(L(t)-y).
\end{equation}
 Both functions $f(y)$ and
$g(y)$ are completely determined by their values at the observed
trajectory positions. These functions are solved by
self-consistently determining $f(\xi_j^{(\alpha)})$ and
$g(\xi_j^{(\alpha)})$ for all $j$ and $\alpha$, which is essentially a
high (though finite)-dimensional root identification problem.

To emphasize this point, and to simplify the root problem, we rewrite
Eq.~\ref{eq:deltafsliced} and Eq.~\ref{eq:dDsliced}, grouping terms by
how they depend on $f$ and $g$. Eq.~\ref{eq:deltafsliced} becomes
{\small
\begin{align} 
0 = f(y)-\frac{1}{2}\sum_{\alpha,j}
G_f(y,\xi_{j}^{(\alpha)})\Bigg[{\xi_{j+1}^{(\alpha)}-\xi_{j}^{(\alpha)}}-\underline{D(\xi_{j}^{(\alpha)})
f(\xi_{j}^{(\alpha)})}\delta t
-\underline{D(\xi_{j}^{(\alpha)})}m(\xi_{j}^{(\alpha)} ,t_j)\delta t -
\underline{D^\prime(\xi_{j}^{(\alpha)})} \delta t\Bigg], \label{eq:froot}
 \end{align}
and Eq.~\ref{eq:dDsliced} becomes 
\begin{align} 
0 &=g(y)+\frac{1}{2}\sum_{\alpha,j}\left[
G_g(y,\xi_j^{(\alpha)})-\frac{\partial
G_g(y,\xi_j^{(\alpha)})}{\partial
z}(\xi_{j+1}^{(\alpha)}-\xi_{j}^{(\alpha)} ) \right] \nonumber\\
 & +\frac{\delta t}{2}\sum_{\alpha,j}\Bigg\{ \underline{D
(\xi_{j}^{(\alpha)})}\Bigg(\frac{\partial
G_g(y,\xi_j^{(\alpha)})}{\partial z}m(\xi_j^{(\alpha)},t_j)+\frac{G_g(y,\xi_j^{(\alpha)})}{2}m^2
(\xi_j^{(\alpha)},t_j) \Bigg)+\underline{D^\prime(\xi_{j}^{(\alpha)})}\left[ \frac{\partial
G_g(y,\xi_j^{(\alpha)})}{\partial z} +
G_g(y,\xi_j^{(\alpha)}) m(\xi_j^{(\alpha)},t_j) \right]\nonumber
\\ & + \underline{D(\xi_{j}^{(\alpha)})f(\xi_{j}^{(\alpha)}) }\left[
\frac{\partial G_g(y,\xi_j^{(\alpha)})}{\partial z} +
G_g(y,\xi_j^{(\alpha)}) m(\xi_j^{(\alpha)},t_j) \right]-
\underline{\frac{1}{D(\xi_{j}^{(\alpha)})}}\frac{G_g(y,\xi_j^{(\alpha)})}{2}
\left (\frac{\xi_{j+1}^{(\alpha)}-\xi_{j}^{(\alpha)} }{\delta t}
\right)^2 \nonumber\\ &+\underline{ g^\prime(\xi_{j}^{(\alpha)})
D^\prime(\xi_{j}^{(\alpha)})}\frac{G_g(y,\xi_j^{(\alpha)})}{2} +
\underline{D(\xi_{j}^{(\alpha)})
f^2(\xi_{j}^{(\alpha)})}\frac{G_g(y,\xi_j^{(\alpha)})}{2}
+ \underline{D^\prime(\xi_{j}^{(\alpha)})
f(\xi_{j}^{(\alpha)}) }G_g(y,\xi_j^{(\alpha)}) \Bigg\}.\label{eq:groot}
\end{align} 
}
\subsection{Approximate solution}
In both Eqs~\ref{eq:froot} and~\ref{eq:groot}, we have underlined the terms which we need
to evaluate.  The size of this problem is two times the number of observed
positions, which in practice is a very large number. Solving this
problem exactly yields a very high resolution recovery of the desired
functions $f$, and $g$, however, since the solution is regularized,
such resolution is unnecessary.  Instead of solving these equations
exactly, we approximate the terms $f(\obsj)$ and $g(\obsj)$ about evenly
spaced control points $y_k$ separated by gaps of length $\delta y\ll
\sqrt{\gamma}$ ($\gamma = \gamma_{f,g}$ are regularization parameters
defining the correlations lengths of $f$ and $g$). Using these points, we approximate
quantities like $f(\xi_j^{(\alpha)})$ by Taylor expansion about the
 nearest $y_k$ to $\xi_j^{(\alpha)}$, and its two nearest neighbors $y_{k-1}$ and $y_{k+1}$ yielding
 the approximation

{\small\begin{align} f(\xi_j^{(\alpha)}) &\approx f(y_{k}) +
(\xi_j^{(\alpha)}-y_k)\left.\frac{df(y)}{\d{y}}\right|_{y_k}+\frac{(\xi_j^{(\alpha)}-y_k)^2}{2}\left.\frac{d^2f(y)}{\d{y}^2}\right|_{y_k}
\nonumber\\ &\approx f(y_{k})
+(\xi_j^{(\alpha)}-y_k)\frac{f(y_{k+1})-f(y_{k-1}) }{2\delta y}
\nonumber\\ &\qquad +
(\xi_j^{(\alpha)}-y_k)^2\frac{f(y_{k+1})-2f(y_k)+f(y_{k-1}) }{2(\delta
y)^2} \label{eq:fapprox}.
\end{align}}
Grouping the terms in Eq.~\ref{eq:fapprox} by $f(y_k)$ yields
\begin{equation}
f(\xi_j^{(\alpha)}) = a_j^{(\alpha)} f(y_{k-1}) +b_j^{(\alpha)} f(y_{k}) +c_j^{(\alpha)} f(y_{k+1})
\end{equation}
where
\begin{equation}
a_j^{(\alpha)} = \left[ \frac{(\xi_j^{(\alpha)}-y_k)^2}{2(\delta y)^2} -\frac{\xi_j^{(\alpha)}-y_k}{2\delta y}\right]
\end{equation}
\begin{equation}
b_j^{(\alpha)} = \left[ 1 -\frac{(\xi_j^{(\alpha)}-y_k)^2}{(\delta y)^2}\right]
\end{equation}
\begin{equation}
c_j^{(\alpha)} = \left[ \frac{(\xi_j^{(\alpha)}-y_k)^2}{2(\delta y)^2} +\frac{\xi_j^{(\alpha)}-y_k}{2\delta y}\right].
\end{equation}
For $D(y)=D_0e^{g(y)}$, we choose to define our approximation directly
on the values $D(y_k)=D_0e^{g(y_k)}$ rather than on Taylor expansions
for $g$:

\begin{align}
D(\xi_j^{(\alpha)}) &\approx a_j^{(\alpha)} D(y_{k-1}) +b_j^{(\alpha)} D(y_{k}) +c_j^{(\alpha)} D(y_{k+1})\nonumber\\
&=D_0\left[a_j^{(\alpha)} e^{g(y_{k-1}) }+b_j^{(\alpha)} e^{g(y_{k}) }+c_j^{(\alpha)} e^{g(y_{k+1})}\right].
\end{align}
We use this approximation because it results in only pairwise products like $f(y_m)D(y_n)$ 
when used in Eqs.~\ref{eq:deltafsliced} and ~\ref{eq:dDsliced} rather than higher 
order terms that would result if one defined $D$ using Taylor expansions in $g$.
Similarly, we will use the approximation for $1/D$,

\begin{align}
\frac{1}{D(\xi_j^{(\alpha)}) }&\approx a_j^{(\alpha)} \frac{1}{D(y_{k-1}) }+b_j^{(\alpha)} \frac{1}{D(y_{k}) }+c_j^{(\alpha)} \frac{1}{D(y_{k+1})}\nonumber\\
&=\frac{1}{D_0}\left[a_j^{(\alpha)} e^{-g(y_{k-1}) }+b_j^{(\alpha)}
e^{-g(y_{k}) }+c_j^{(\alpha)} e^{-g(y_{k+1})}\right].
\end{align}

With these substitutions in place, one may evaluate
Eqs.~\ref{eq:deltafsliced}-\ref{eq:dDsliced} given values $f(y_k),$
$g(y_k),$ $g^\prime(y_k),$ $D(y_k),$ $1/D(y_k)$.  The coefficients in front
of each of these terms is data dependent and need only be evaluated
a single time for a given choice of control points.
We also approximate the kernel values like $G_f(\obsj,\obs{k}{\alpha})$ by
evaluating the kernels about the nearest control points.
The resulting root problem of Eqs.~\ref{eq:froot},~\ref{eq:groot}
is solved using \texttt{scipy.optimize.root} in our implementation available at
\url{https://github.com/joshchang/dfsinference}.

\section{Semiclassical approximation}

We will  denote the partial derivative of a kernel with 
respect to its left coordinate as $\partial_y$, and with respect to the right coordinate 
as $\partial_z$. 
To construct the semiclassical approximation to the Hamiltonian, one
needs to evaluate the second variational derivatives. We begin with
the Hessian of the Hamiltonian with respect to $f$, 

{\small
\begin{equation}
\ddH{f(y)}{f(z)} = \Bigg[R_f(-\Delta)+\frac{1}{2}\sum_{\alpha,j} \delta(z-\xi_j^{(\alpha)})D(z)\delta t\Bigg]\delta(y-z).
\label{eq:Hff}
\end{equation}
}
We wish to compute the operator inverse
\begin{equation}
{H}_{ff}(y,z) \equiv  \left[\ddH{f(y)}{f(z)} \right]^{-1}
\end{equation}
which obeys the relationship
\[
\int \ddH{f(y)}{f(x)}{H}_{ff}(x,z)dx = \delta(y-z).
\]
Applying this relationship, and convolving both sides by the Greens
function $G_f$ for $R_f$ yields

\begin{equation}
{H}_{ff}(y,z)= G_f(y,z)  - \frac{\delta t}{2}\sum_{j,\alpha} \overbrace{G_f(y,\xi_j^{(\alpha)})D(\xi_j^{(\alpha)})}^{\textrm{known}} \overbrace{({H}_{ff}(y,z))_{y=\xi_j^{(\alpha)}}}^{\textrm{unknown}}.\label{eq:Hffyz1}
\end{equation}
Eq.~\ref{eq:Hffyz1} can be determined analytically by solving an equivalent
linear system for the unknown term in the sum. In practice, the
solution of this system is prohibitive due to large size. In the
same spirit as in inference, we approximate the inversion using
function evaluations interpolated about the same control points $y_k$ that we have used
before. For the sake of simplicity, we will utilize a leading-order approximation
for each of the unknown functions as opposed to the higher-order scheme that we used
for inference.

 Our problem is then transformed into the smaller problem of solving for 
each control point the equation
\begin{equation}
{H}_{ff}(y_m,z) \approx G_f(y_m,z) - \sum_{k} G_f(y_m,y_k)n_kD(y_k) 
\overbrace{{H}_{ff}(y_k,z)}^{\textrm{unknown}}\label{eq:Hffyk},
\end{equation}
where $n_k$ is the number of trajectory positions that are nearest to $y_k$. Eq.~\ref{eq:Hffyk} 
has a solution that can be represented as

\begin{equation}
\mathbf{H}_{ff} = (\mathbf{I}+\mathbf{d})^{-1} \mathbf{M}^{-1}_f
\end{equation}
where $({\mathbf{H}}_{ff})_{mn} = \mathbf{H}_{ff}(y_m,y_n) $ is a
matrix of values on the left hand side of Eq.~\ref{eq:Hffyk},
$\mathbf{M}_f$ is a matrix of values
$(\mathbf{M}^{-1})_{mn}=G_f(y_m,y_n)$, and $\mathbf{d}$ is a matrix
of values $(G(y_k,y_m)n_mD(_m))_{km}$.

We undertake the same procedure for the Hessian with respect to $g$.
After some algebra, we find that

\begin{align}
 \ddH{g(y)}{g(z)} &=R_g(-\Delta) \delta(y-z) + \frac{1}{2}\sum_{j,\alpha}\delta(y-z)\delta(z-\xi_j^{(\alpha)}) \left[\frac{(\xi_{j+1}^{(\alpha)}-\xi_j^{(\alpha)})^2+A(z,t_j)^2\delta t^2  }{2D(z)\delta t} \right]  \nonumber\\
&\quad-\delta(y-z)\frac{1}{2}\frac{\partial}{\partial z}\left[\sum_{j,\alpha}\delta(z-\xi_j^{(\alpha)})A(z,t_j)\right]\delta t-\frac{1}{2}\frac{\partial}{\partial z}\left[D(z)\frac{\partial\delta(y-z)}{\partial z}\delta(z-\xi_j^{(\alpha)})\right]\delta t.\label{eq:Hgg}
\end{align}

Inversion of this operator is slightly more involved than the previous
operator due to the presence of derivatives. Let us write

\begin{equation}
H_{gg}(y,z)\equiv\left[ \ddH{g(y)}{g(z)} \right]^{-1}.
\end{equation}
After convolving an appropriate Greens function $G_g$, the
inverse operator satisfies the relationship
\begin{align}
\lefteqn{H_{gg}(y,z) =  G_g(y,z)}\nonumber\\
&\quad -\sum_{j,\alpha}\left\{ G_g(y,\xi_j^{(\alpha)})\frac{(\xi_{j+1}^{(\alpha)}-\xi_j^{(\alpha)})^2+A(\xi_j^{(\alpha)},t_j)^2\delta t^2}{4D(\xi_j^{(\alpha)})\delta t}+\frac{\delta t\partial_z G_g(y,\xi_j^{(\alpha)}) A(\xi_j^{(\alpha)}, t_j)}{2}\right\}H_{gg}(\xi_j^{(\alpha)},z)\nonumber\\
&\qquad -\frac{\delta t}{2} \sum_{j,\alpha}\left\{ G_g(y,\xi_j^{(\alpha)}) A (\xi_j^{(\alpha)},t_j) +\partial_z G_g(y,\xi_j^{(\alpha)}) D (\xi_j^{(\alpha)})\right\}\left[\frac{\partial H_{gg}(\xi_j^{(\alpha)},z)}{\partial y}\right].  \label{eq:Lambda1}
\end{align}
It is evident that $H_{gg}$ is known self-consistently if
$H_{gg}(\xi_j^{(\alpha)},z)$, and
$\partial_yH_{gg}(\xi_j^{(\alpha)},z)$ are all known. Differentiating
Eq.~\ref{eq:Lambda1}, one finds 

\begin{align}
\lefteqn{ \partial_y H_{gg}(y,z) = \partial_y  G_g(y,z)}\nonumber\\
& -\sum_{j,\alpha}\left\{ \partial_y G_g(y,\xi_j^{(\alpha)})\frac{(\xi_{j+1}^{(\alpha)}-\xi_j^{(\alpha)})^2+A(\xi_j^{(\alpha)},t_j)^2\delta t^2}{4D(\xi_j^{(\alpha)})\delta t}+\frac{\delta t\partial_y \partial_z G_g(y,\xi_j^{(\alpha)}) A(\xi_j^{(\alpha)}, t_j)}{2}\right\}H_{gg}(\xi_j^{(\alpha)},z)\nonumber\\
&\qquad -\frac{\delta t}{2} \sum_{j,\alpha}\left\{\partial_y  G_g(y,\xi_j^{(\alpha)}) A (\xi_j^{(\alpha)},t_j) +\partial_y \partial_z G_g(y,\xi_j^{(\alpha)}) D (\xi_j^{(\alpha)})\right\}\left[\frac{\partial H_{gg}(\xi_j^{(\alpha)},z)}{\partial y}\right].
\label{eq:Lambda2}
\end{align}
Eqs.~\ref{eq:Lambda1} and~\ref{eq:Lambda2} can be solved 
together at the control points by solving an associated linear system
\begin{align}
\boldsymbol\Lambda_1 &=\mathbf{M}_1- \mathbf{A}_1\boldsymbol\Lambda_1- \mathbf{A}_2\boldsymbol\Lambda_2 \label{eq:linearLam1} \\
\boldsymbol\Lambda_2 &=\mathbf{M}_2- \mathbf{A}_3\boldsymbol\Lambda_1- \mathbf{A}_4\boldsymbol\Lambda_2  \label{eq:linearLam2}
\end{align}
where the vectors $\boldsymbol\Lambda_1, \boldsymbol\Lambda_2$ contain
entries $H_{gg}(y_m,y_n)$ and $\partial_yH_{gg}(y_m,y_n)$
respectively. The vectors $\mathbf{M}_1,\mathbf{M}_2$ contain entries
$G_g(y_m,y_n)$ and $\partial_yG_g(y_m,y_n)$ respectively,
and all of the $\mathbf{A}_{(\cdot)}$ terms are matrices.

Finally, we have the mixed term
{\small
\begin{align}
\frac{\delta H}{\delta f(y)\delta g(z)} &= \frac{1}{2}
\sum_{\alpha,j}\delta(y-z) \delta(z-\xi_j^{(\alpha)})\Bigg[D^\prime(z) + D(z)\big[f(z)+m(z,t) \big] \Bigg]\delta t \nonumber\\
&\qquad\qquad-\frac{1}{2}\sum_{\alpha,j} \frac{\partial}{\partial z}
\left[\delta(y-z)\delta(z-\xi_j^{(\alpha)})D(z) \right] \delta t
\label{eq:Hfg}.
\end{align}}

Using these expressions, we can approximate the semiclassical
posterior variance in both $f$ and $g$.  For $f$, we have

\begin{align}
\Sigma_{ff}&\equiv\big\langle \left(f(y)-f^\star(y)\right)  
\left(f(z)-f^\star(z) \right)  \big\rangle\nonumber\\
& =\left[\ddH{f(y)}{f(z)}- \ddH{f(y)}{g(z)} \left(\ddH{g(y)}{g(z)} \right)^{-1}
\frac{\delta^2 H}{\delta g(y)\delta f(z)}  \right]^{-1} \nonumber\\
%&=\left[\ddH{f(y)}{f(z)}\right]^{-1}\Bigg\{ \mathbf{I} \nonumber\\
%&+ \ddH{f(y)}{f(z)}\left[\left( \ddH{g(y)}{g(z)}\right)^{-1}+\ddH{g(y)}{f(z)}\left(\ddH{f(y)}{f(z)}\right)^{-1}\ddH{g(y)}{f(z)} \right]^{-1}\ddH{g(y)}{f(z)}\left( \ddH{g(y)}{g(z)}\right)^{-1}  \Bigg\}
\end{align}
and similarly an estimate for $g$
\begin{align}
\Sigma_{gg}&\equiv\big\langle \left(g(y)-g^\star(y)\right)  
\left(g(z)-g^\star(z) \right)  \big\rangle \nonumber\\
& \left[\ddH{g(y)}{g(z)}- \ddH{g(y)}{f(z)} 
\left(\ddH{f(y)}{f(z)}\right)^{-1}\ddH{f(y)}{g(z)}  \right]^{-1}.
\end{align}
From these expressions, it is evident that the recovery errors of $f$
and $g$ are coupled. Given the error for $g$, one can approximate the
pointwise error in the recovery of $D(y)$ as

\begin{align}
\lefteqn{\langle D^2(y)\rangle - \langle D(y)\rangle^2 = \langle D_0^2e^{2g(y)}\rangle-\langle D_0e^{g(y)}\rangle^2} \nonumber\\
&= \sum_{n=0}^\infty\frac{\langle D_0^2 2^ng^n(y)\rangle}{n!}-\left[  \sum_{n=0}^\infty\frac{\langle D_0g^n(y)\rangle}{n!}\right]^2\nonumber\\
&\sim D_0^2\langle g^2(x)\rangle.
\end{align}
The expectation values with respect to $g$ can be computed to
higher orders using Feynman diagrams.

%nuum limit in the sense of left--Riemann integration.

\section{Posterior covariances}

Our goal is to compute $\Sigma_{ff}$ and $\Sigma_{gg}$ which
will involve terms which we have computed via Eqs.~\ref{eq:Hgg},~\ref{eq:Hff}.

\subsection{Posterior covariance of $f$}
For $\Sigma_{ff}$, 
\begin{align*}
\Sigma_{ff}(y,z)=\left[\ddH{f(y)}{f(z)}-\ddH{f(y)}{g(z)}\left(\ddH{g(y)}{g(z)}\right)^{-1}\ddH{g(y)}{f(z)} \right]^{-1}
\end{align*}
where
\begin{align*}
\ddH{f(y)}{g(z)} = \frac{1}{2}\delta(y-z)\sum_{j,\alpha}\delta(z-\obsj)A(z,t_j)\delta t -\frac{1}{2}\frac{\partial}{\partial z}\left[\delta(y-z)\sum_{j,\alpha}\delta(z-\obsj) D(z)\right]\delta t
\end{align*}
and its adjoint is
\begin{align*}
\ddH{g(y)}{f(z)} = \frac{1}{2}\delta(y-z)\sum_{j,\alpha}\delta(z-\obsj)A(z,t_j)\delta t +\frac{1}{2}\frac{\partial}{\partial z}\delta(y-z)\sum_{j,\alpha}\delta(z-\obsj) D(z)\delta t.
\end{align*}
Recalling that
\[
H_{gg}(y,z) \equiv \left(\ddH{g(y)}{g(z)}\right)^{-1}.
\]

We compute first
\begin{align}
\lefteqn{\ddH{f(y)}{g(z)}\left(\ddH{g(y)}{g(z)}\right)^{-1} }\nonumber\\
&=\int \left\{\frac{1}{2}\delta(y-z)\sum_{j,\alpha}\delta(z-\obsj)A(z,t_j)\delta t -\frac{1}{2}\frac{\partial}{\partial z}\left[\delta(y-z)\sum_{j,\alpha}\delta(z-\obsj) D(z)\right]\delta t \right\}H_{gg}(z, u) \d{z} \nonumber\\
&=\left(\frac{\delta t}{2}\right)\left\{  \sum_{j,\alpha}\delta(y-\obsj)A(y,t_j)H_{gg}(y,u)+  \sum_{j,\alpha}\delta(y-\obsj)D(y)\partial_y H_{gg}(y,u)\right\} \nonumber\\
&=\frac{\delta t}{2}\sum_{j,\alpha}\delta(y-\obsj)\left[A(y,t_j)H_{gg}(y,u)+D(y)\partial_y H_{gg}(y,u)\right].
\end{align}
Now we can compute
\begin{align}
\lefteqn{\ddH{f(y)}{g(z)}\left(\ddH{g(y)}{g(z)}\right)^{-1}\ddH{g(y)}{f(z)} }\nonumber\\
&=\int \frac{\delta t}{2}\sum_{j,\alpha}\delta(y-\obsj)\left[A(y,t_j)H_{gg}(y,u)+D(y)\partial_y H_{gg}(y,u)\right] \nonumber\\
&\quad\times \left[\frac{1}{2}\delta(u-z)\sum_{j,\alpha}\delta(z-\obsj)A(z,t_j)\delta t +\frac{1}{2}\frac{\partial}{\partial z}\delta(u-z)\sum_{k,\beta}\delta(z-\obs{k}{\beta}) D(z)\delta t \right] \d{u}\nonumber\\
&= \left(\frac{\delta t}{2}\right)^2 \sum_{j,\alpha}\delta(y-\obsj)\sum_{k,\beta}\delta(z-\obs{k}{\beta})  \left[ A(y,t_j)H_{gg}(y,z)+D(y)\partial_y H_{gg}(y,z) \right]A(z,t_k) \nonumber\\
&\quad + \left(\frac{\delta t}{2}\right)^2 \sum_{j,\alpha}\delta(y-\obsj)\sum_{k,\beta}\delta(z-\obs{k}{\beta})  \left[ A(y,t_j)\partial_zH_{gg}(y,z)+D(y)\partial_z\partial_y H_{gg}(y,z) \right] D(z) .
\end{align}

Now we have

\begin{align}
\lefteqn{\Sigma_{ff}(y,z) = G_f(y,z)-\frac{\delta t}{2}\sum_{j,\alpha} G_f(y,\obsj)D(\obsj)\Sigma_{ff}(\obsj,z)} \nonumber\\
&+ \left(\frac{\delta t}{2}\right)^2 \sum_{j,\alpha}G_f(y,\obsj)  \sum_{k,\beta}  \left[ A(\obsj,t_j)H_{gg}(\obsj,\obs{k}{\beta})+D(\obsj)\partial_y H_{gg}(\obsj,\obs{k}{\beta}) \right]A(\obs{k}{\beta},t_k)\Sigma_{ff}(\obs{k}{
\alpha},z) \nonumber \\ 
&+ \left(\frac{\delta t}{2}\right)^2 \sum_{j,\alpha}G_f(y,\obsj)  \sum_{k,\beta}  \left[ A(\obsj,t_j)\partial_zH_{gg}(\obsj,\obs{k}{\beta})+D(\obsj)\partial_z\partial_y H_{gg}(\obsj,\obs{k}{\beta}) \right]D(\obs{k}{\beta})\Sigma_{ff}(\obs{k}{
\alpha},z). \label{eq:Sigmaff}
\end{align}
This equation can be solved in the same manner as Eq.~\ref{eq:Lambda2} by solving a linear system similar to that found in Eq.~\ref{eq:Hffyk}.
\subsection{Posterior covariance of $g$}
For computing $\Sigma_{gg}$, we use compute the operator inverse
\[\Sigma_{gg}(y,z)= \left[\ddH{g(y)}{g(z)}- \ddH{g(y)}{f(z)} \left( \ddH{f(y)}{f(z)}\right)^{-1}\ddH{f(y)}{g(z)}  \right]^{-1}. \]

Recalling that 
\[
H_{ff}(y,z) \equiv \left(\ddH{f(y)}{f(z)}\right)^{-1}.
\]
\begin{align}
\lefteqn{\ddH{g(y)}{f(z)}\left(\ddH{f(y)}{f(z)}\right)^{-1} }\nonumber\\
&=\int \left[\frac{1}{2}\delta(y-z)\sum_{j,\alpha}\delta(z-\obsj)A(z,t_j)\delta t +\frac{1}{2}\frac{\partial}{\partial z}\delta(y-z)\sum_{j,\alpha}\delta(z-\obsj) D(z)\delta t \right]H_{ff}(z,u) \d{z} \\
&=\frac{\delta t }{2}\sum_{j,\alpha}\left[\delta(y-\obsj)A(y,t_j)H_{ff}((y,u)-\frac{\partial}{\partial y} \left( \delta(y-\obsj)D(y)H_{ff}(y,u) \right)\right].
\end{align}

\begin{align}
\lefteqn{\ddH{g(y)}{f(z)}\left(\ddH{f(y)}{f(z)}\right)^{-1}\ddH{f(y)}{g(z)} }\nonumber\\
&=\left( \frac{\delta t }{2}\right)^2\int \sum_{j,\alpha}\left[\delta(y-\obsj)A(y,t_j)H_{ff}(y,u)-\frac{\partial}{\partial y} \left( \delta(y-\obsj)D(y)H_{ff}(y,u) \right)\right] \nonumber\\
&\qquad\times\left\{\delta(u-z)\sum_{k,\beta}\delta(z-\obs{k}{\beta})A(z,t_k)-\frac{\partial}{\partial z}\left[\delta(u-z)\sum_{k,\beta}\delta(z-\obs{k}{\beta}) D(z)\right]\right\} \d{u} \nonumber\\
&=\left( \frac{\delta t }{2}\right)^2\sum_{j,\alpha}\left[\delta(y-\obsj)A(y,t_j)H_{ff}(y,z)-\frac{\partial}{\partial y} \left( \delta(y-\obsj)D(y)H_{ff}(y,z) \right)\right]\sum_{k,\beta}\delta(z-\obs{k}{\beta})A(z,t_k) \nonumber\\
&\quad-\left( \frac{\delta t }{2}\right)^2\sum_{j,\alpha}\delta(y-\obsj)A(y,t_j) \partial_z(H_{ff}(y,z)\sum_{k,\beta}\delta(z-\obs{k}{\beta}) D(z)) \nonumber\\
&\quad+\left( \frac{\delta t }{2}\right)^2 \frac{\partial}{\partial y}\frac{\partial}{\partial z}\left[ \sum_{j,\alpha}\delta(y-\obsj)D(y)H_{ff}(y,z) \sum_{k,\beta}\delta(z-\obs{k}{\beta}) D(z)\right].
\end{align}

\begin{align}
\lefteqn{\Sigma_{gg}(y,z) = G_g(y,z) }\nonumber\\
& -\sum_{j,\alpha} G_g(y,\xi_j^{(\alpha)})\frac{(\xi_{j+1}^{(\alpha)}-\xi_j^{(\alpha)})^2+A(\xi_j^{(\alpha)},t_j)^2\delta t^2}{4D(\xi_j^{(\alpha)})\delta t}\Sigma_{gg}(\xi_j^{(\alpha)},z)-\frac{1}{2}  \sum_{j,\alpha}\partial_z G_g(y,\xi_j^{(\alpha)}) A(\xi_j^{(\alpha)}, t_j)\Sigma_{gg}(\xi_j^{(\alpha)},z)\delta t\nonumber\\
& -\frac{1}{2} \sum_{j,\alpha} G_g(y,\xi_j^{(\alpha)}) A (\xi_j^{(\alpha)},t_j)\left[\frac{\partial \Sigma_{gg}(\xi_j^{(\alpha)},z)}{\partial y}\right]\delta t  -\frac{1}{2} \sum_{j,\alpha} \partial_z G_g(y,\xi_j^{(\alpha)}) D (\xi_j^{(\alpha)})\left[\frac{\partial \Sigma_{gg}(\xi_j^{(\alpha)},z)}{\partial y}\right]\delta t \nonumber \\
&+  \left( \frac{\delta t }{2}\right)^2\sum_{j,\alpha}\sum_{k,\beta}\left[G_g(y,\obsj)A(\obsj,t_j)H_{ff}(\obsj,\obs{k}{\beta}) +\partial_zG_g(y,\obsj) D(\obsj)H_{ff}(\obsj,\obs{k}{\beta}) \right]A(\obs{k}{\beta},t_k) \Sigma_{gg}(\obs{k}{\beta},z) \nonumber \\
&\quad+\left( \frac{\delta t }{2}\right)^2\sum_{j,\alpha}\sum_{k,\beta}G_g(y,\obsj)A(\obsj,t_j) H_{ff}(\obsj,\obs{k}{\beta})D(\obs{k}{\beta})\partial_y\Sigma_{gg}(\obs{k}{\beta},z) \nonumber\\
&\quad+\left( \frac{\delta t }{2}\right)^2  \sum_{j,\alpha}\sum_{k,\beta}\partial_zG_g(y,\obsj)D(\obsj)H_{ff}(\obsj,\obs{k}{\beta}) D(\obs{k}{\beta}) \partial_y\Sigma_{gg}(\obs{k}{\beta},z) .
\label{eq:Sigmagg}
\end{align}

This equation can be solved in the same manner as Eq.~\ref{eq:Lambda2} by solving a linear system similar to that found in Eqs.~\ref{eq:linearLam1},~\ref{eq:linearLam2}.

\section{Computing the marginal likelihood}

To compute the marginal likelihood function (Eq.~\ref{eq:logmarginal}) given a choice of regularization parameters, there are several quantities that need to be computed. Here we give detailed computations for each of these quantities. First, we need to evaluate the Hamiltonian at the saddle solutions. This calculation requires first the norms
\begin{align}
\int f^* R_f(-\Delta) f^* \d{y} &= \int f^*(x)  \frac{1}{2}\sum_{\alpha,j} \delta(y-\xi_j^\alpha) \left[ \xi_{j+1}^{(\alpha)}-\xi_j^{(\alpha)}-A(y,t)\delta t \right] \d{y} \nonumber\\
&=  \frac{1}{2}\sum_{\alpha,j} f^*(\xi_j^{(\alpha)})\left[ \xi_{j+1}^{(\alpha)}-\xi_j^{(\alpha)}-A(\xi_j^{(\alpha)} ,t)\delta t \right]
\end{align}
and
\begin{align}
\lefteqn{\int g^* R_g(-\Delta) g^* \d{y} = -\frac{1}{2}\int g^*\Bigg\{\sum_{\alpha,j}\frac{\partial}{\partial y}\left[ \delta(y-\xi_{j}^{(\alpha)} ) (\xi_{j+1}^{(\alpha)} -\xi_j^{(\alpha)} -A( \xi_j^{(\alpha)}  ,t_j)\delta t)  \right]   }\nonumber\\
&\qquad\qquad\qquad+\sum_{\alpha,j} \delta(y-\xi_j^{(\alpha)})\Bigg[1-\frac{ (\xi_{j+1}^{(\alpha)} -\xi_j^{(\alpha)})^2 - A^2(y,t_j) (\delta t)^2 }{2D(y) \delta t} \Bigg] \Bigg\} \d{y} \nonumber \\
&\qquad=\frac{1}{2}\sum_{\alpha,j} \left[g^{*\prime}(\xi_j^{(\alpha)})(\xi_{j+1}^{(\alpha)} -\xi_j^{(\alpha)} -A( \xi_j^{(\alpha)}  ,t_j)\delta t)   \right]- \frac{1}{2}\sum_{\alpha,j}g^*(\xi_j^{(\alpha)})\Bigg[1-\frac{ (\xi_{j+1}^{(\alpha)} -\xi_j^{(\alpha)})^2 - A^2(\xi_j^{(\alpha)},t_j) (\delta t)^2 }{2D(\xi_j^{(\alpha)}) \delta t} \Bigg].
\end{align}

Altogether, the Hamiltonian portion of the marginal likelihood is
\begin{align*}
\lefteqn{H[f^*,g^*] = \frac{1}{4}\sum_{\alpha,j} f^*(\xi_j^{(\alpha)})\left[ \xi_{j+1}^{(\alpha)}-\xi_j^{(\alpha)}-A(\xi_j^{(\alpha)} ,t)\delta t \right]}\nonumber\\
&\qquad+\frac{1}{4}\sum_{\alpha,j} \left[g^{*\prime}(\xi_j^{(\alpha)})(\xi_{j+1}^{(\alpha)} -\xi_j^{(\alpha)} -A( \xi_j^{(\alpha)}  ,t_j)\delta t)   \right]\nonumber\\
&\qquad- \frac{1}{4}\sum_{\alpha,j}g^*(\xi_j^{(\alpha)})\Bigg[1-\frac{ (\xi_{j+1}^{(\alpha)} -\xi_j^{(\alpha)})^2 - A^2(\xi_j^{(\alpha)},t_j) (\delta t)^2 }{2D^*(\xi_j^{(\alpha)}) \delta t} \Bigg]+\frac{1}{2}\sum_{\alpha,j} \log D^*(\xi_j^{(\alpha)}) \nonumber\\
&\quad + \sum_{\alpha,j}\frac{{\left(\xi^{(\alpha)}_{j+1}-\xi^{(\alpha)}_j-A(\xi_j^{(\alpha)},t_j)\delta t\right)^2} }{4D^*(\xi_j^{(\alpha)})\delta t}.
\end{align*}

The other component of the marginal likelihood is the term
\begin{equation}
\Tr\log\boldsymbol{\Sigma}-\Tr\log G_f - \Tr\log G_g.
\end{equation}
We note now that this expression is equivalent to 
\begin{equation}
-\Tr\log\boldsymbol{\Sigma}^{-1}-\Tr\log G_f - \Tr\log G_g = -\log\det \Sigma_{ff}^{-1}G_f-\log\det \Sigma_{gg}^{-1} G_g.
\end{equation}
 $\Sigma_{ff}^{-1}$ is the upper left quadrant of the semiclassical Hessian matrix, and $\Sigma_{gg}^{-1}$ is the lower right quadrant of the Hessian matrix. These determinants can be calculated exactly through the solution of an eigenvalue problem.
First we compute the (right) eigenfunctions $\varphi_n$ and eigenvalues $\lambda_n$ of the operator 
\begin{equation}
\Sigma_{ff}^{-1}G_f = \delta(y-z) + \frac{\delta t}{2}\sum_{j,\alpha}G(y,z)D(y)\delta(y-\obsj).
\end{equation}
 They  satisfy the relationship
\begin{equation}
(1-\lambda_n)\varphi_n(z) + \frac{\delta t}{2}\sum_{j,\alpha} D(\obsj)G_f(\xi_j^{(\alpha)},z) \varphi_n(\xi_j^{(\alpha)})  = 0.
\end{equation}
Plugging in each of the $\xi_j^{(\alpha)}$ in for $z$ yields the condition
\begin{equation}
\left[  \mathbf{G}_D -(\lambda_n-1)\mathbb{I}  \right] \boldsymbol{\varphi} = 0,\label{eq:eig1}
\end{equation}
where $\boldsymbol{\varphi}=[\varphi_n(\xi_1^{(1)} )\quad \varphi_n(\xi_2^{(1)})\quad \ldots]$ is a vector and $\mathbf{G}_D$ is a matrix
\begin{equation}
\mathbf{G}_D = \frac{\delta t}{2} \left(\begin{matrix}
D(\xi_1^{(1)})G_f(\xi_1^{(1)},\xi_1^{(1)}) &  D(\xi_2^{(1)})G_f(\xi_1^{(1)},\xi_1^{(2)}) & \cdots & D(\xi_N^{(M)})G_f(\xi_1^{(1)},\xi_N^{(M)})  \\
D(\xi_1^{(1)})G_f(\xi_2^{(1)},\xi_1^{(1)}) & D(\xi_2^{(1)})G_f(\xi_2^{(1)},\xi_2^{(2)})& \cdots & D(\xi_N^{(M)})G_f(\xi_2^{(1)},\xi_N^{(M)})  \\
D(\xi_1^{(1)})G_f(\xi_3^{(1)},\xi_1^{(1)}) & \cdots \\
D(\xi_1^{(1)})G_f(\xi_4^{(1)},\xi_1^{(1)}) & \cdots \\
\vdots \\
\end{matrix}  \right)
\end{equation}
From Eq.~\ref{eq:eig1}, it is evident that the eigenvalues of the operator $\Sigma_{ff}^{-1}G_g$ can be computed by taking the eigenvalues of $\mathbf{G}_D$ and adding one to each of them. Knowing the eigenvalues of the operator, we have
\begin{equation}
 \log\det \Sigma_{ff}^{-1}G_f  = \sum_n\log \lambda_n.
 \end{equation} 

We do the same for the other operator $\Sigma^{-1}_{ff}G_g$.
We proceed as before. First we compute the operator 
\begin{align}
\Sigma^{-1}_{ff}G_g &= \int \Bigg\{  R_g(-\Delta) \delta(y-z) + \frac{1}{2}\sum_{j,\alpha}\delta(y-z)\delta(z-\xi_j^{(\alpha)}) \left[\frac{(\xi_{j+1}^{(\alpha)}-\xi_j^{(\alpha)})^2+A(z,t_j)^2\delta t^2  }{2D(z)\delta t} \right]  \nonumber\\
&\quad-\delta(y-z)\frac{1}{2}\frac{\partial}{\partial z}\left[\sum_{j,\alpha}\delta(z-\xi_j^{(\alpha)})A(z,t_j)\right]\delta t-\frac{1}{2}\frac{\partial}{\partial z}\left[D(z)\frac{\partial\delta(y-z)}{\partial z}\delta(z-\xi_j^{(\alpha)})\right]\delta t \Bigg\}G_g(z,u) \d{u}\nonumber\\
&= \delta(y-u)+\frac{1}{2}\sum_{j,\alpha} G(y,u)\delta(y-\obsj)  \left[\frac{(\xi_{j+1}^{(\alpha)}-\xi_j^{(\alpha)})^2+A(y,t_j)^2\delta t^2  }{2D(y)\delta t} \right] \nonumber\\
&\quad-\frac{\delta t}{2}\sum_{j,\alpha} G(y,u)\frac{\partial}{\partial y}\left[ \delta(y-\obsj) A(y,t_j)\right]-\frac{\delta t}{2}\sum_{j,\alpha}\frac{\partial}{\partial y}\left[ D(y)\frac{\partial G(y,u)}{\partial y}\delta(y-\obsj)\right].
\end{align}
This operator has left eigenfunctions $\phi_n$ corresponding to eigenvalues $\nu_n$ satisfying the relationship
\begin{align}
(\nu_n-1)\phi_n(z) &=\frac{1}{2}\sum_{j,\alpha}G_g(\obsj,z) \left[\frac{(\xi_{j+1}^{(\alpha)}-\xi_j^{(\alpha)})^2+A(\obsj,t_j)^2\delta t^2  }{2D(\obsj)\delta t} \right]\phi_n(\obsj) \nonumber\\
&+\frac{\delta t}{2}\sum_{j,\alpha}\partial_yG_g(\obsj,z) A(\obsj,t_j)\phi_n(\obsj) \nonumber \\
&+\frac{\delta t}{2}\sum_{j,\alpha}G_g(\obsj,z) A(\obsj,t_j)\phi_n^\prime(\obsj) +\frac{\delta t}{2}\sum_{j,\alpha}\partial_yG_g(\obsj,z)D(\obsj)\phi^\prime_n(\obsj).\label{eq:eigeng1}
\end{align}
The eigenfunctions are determined by their values and derivatives at the observed positions $\obsj$. The derivatives of the eigenfunctions
satisfy the relationship
 \begin{align}
(\nu_n-1)\phi_n^\prime(z) &=\frac{1}{2}\sum_{j,\alpha}\partial_zG_g(\obsj,z) \left[\frac{(\xi_{j+1}^{(\alpha)}-\xi_j^{(\alpha)})^2+A(\obsj,t_j)^2\delta t^2  }{2D(\obsj)\delta t} \right]\phi_n(\obsj) \nonumber\\
&+\frac{\delta t}{2}\sum_{j,\alpha}\partial_z\partial_yG_g(\obsj,z) A(\obsj,t_j)\phi_n(\obsj) \nonumber \\
&+\frac{\delta t}{2}\sum_{j,\alpha}\partial_zG_g(\obsj,z) A(\obsj,t_j)\phi_n^\prime(\obsj) +\frac{\delta t}{2}\sum_{j,\alpha}\partial_z\partial_yG_g(\obsj,z)D(\obsj)\phi^\prime_n(\obsj).\label{eq:eigeng2}
\end{align}
Eqs.~\ref{eq:eigeng1} and~\ref{eq:eigeng2}, can be solved by solving them simultaneously for each $\obsj$. This
solution is found by solving the linear system
\begin{align}
(\nu_n-1)\left[\begin{matrix}\boldsymbol{\phi}_n & \boldsymbol{\phi}^\prime_n \end{matrix}\right] =\left[\begin{matrix}\boldsymbol{\phi}_n & \boldsymbol{\phi}^\prime_n \end{matrix}\right] \left[\begin{matrix} \boldsymbol{A} & \boldsymbol{B} \\ \boldsymbol{C} & \boldsymbol{D}\end{matrix} \right],
\end{align}
where $\boldsymbol{\phi}_n =[\ldots,\phi_n(\obsj),\ldots]$, $\boldsymbol{\phi}^\prime_n = [\ldots,\phi_n(\obsj),\ldots]$, $\mathbf{A}$ is a matrix representing the terms that multiply $\phi_n$ in Eq.~\ref{eq:eigeng1}, $\mathbf{B}$ is a matrix of terms that multiply $\phi^\prime_n$ in Eq.~\ref{eq:eigeng1}, $\mathbf{C}$ is a matrix of terms that multiply $\phi_n$ in Eq.~\ref{eq:eigeng2}, and $\mathbf{D}$ is a matrix of terms that multiply $\phi^\prime_n$ in Eq.~\ref{eq:eigeng2}. It is evident that the eigenvalues $\nu_n$ are simply the eigenvalues of the matrices $\mathbf{A}$ and $\mathbf{D}$, plus one. One then may proceed to minimize Eq.~\ref{eq:logmarginal} through a search algorithm, for instance through the usage
of the Python package \texttt{hyperopt}.

\section{Analysis of regularization}

The estimation of
$D_{0}^{\star}$, the functions $f^{\star}(x)$ and $g^{\star}(x)$, and the
appropriate regularization parameters all hinge on a sufficient number
of trajectory measurements.  Related to the question of uncertainty
quantification is the question how the experiments should be pulled in
order to most-efficiently yield a precise reconstruction of the bond force and
diffusivity.

To examine these issues, we consider the semiclassical Hessian matrix
$\mathbf{\Sigma}^{-1}$ in the situation where we wish to estimate the functions
$f$ and $g$ at a position $y$, given $n$ trajectory position measurements taken at a single
position $x$; i.e, we are assuming that we are observing $n$ independent trajectory
displacements $\{d_j\}^n_{j=1}$ originating from $x$. The $j$ index will be used to identify the incidental force applied by the pulling apparatus. For this situation, we can rewrite the Hessian matrix
\begin{equation}
\boldsymbol{\Sigma}^{-1}(y,z) = \left[\begin{matrix} R_f(-\Delta)\delta(y-z)+\frac{n \delta t}{2}\delta(z-x)\delta(y-z) & \begin{smallmatrix} \frac{n\delta t}{2}\delta(y-z)\delta(z-x)A(z,t_j) \\ -\frac{n\delta t}{2} \frac{\partial}{\partial z}[ \delta(y-z)\delta(z-x)D(z)]  \end{smallmatrix}  \\ \\
\begin{smallmatrix}\frac{n \delta t}{2} \delta(y-z)\delta(z-x)A(z,t_j)\\ +\frac{n \delta t}{2}\frac{\partial}{\partial z}\delta(y-z) \delta(z-x)D(z) \end{smallmatrix} & \begin{smallmatrix} R_g(-\Delta)\delta(y-z) +\frac{n\delta t}{2}\delta(y-z)\delta(z-x)\left[\frac{d_j^2 + A(z,t_j)^2\delta t^2}{2D(z)\delta t} \right] \\  -\frac{n\delta t}{2}\left[\delta(y-z)\partial_z( \delta(z-x)A(z,t_j))+\partial_z(D(z)\partial_z\delta(y-z)\delta(z-x)) \right]\end{smallmatrix}
 \end{matrix}\right]
\end{equation}
and use it to approximate the posterior variance in the estimator in the large-$n$ limit. The inverse of the upper right quadrant of this matrix can be computed by solving a system of two equations for $H_{ff}(y,z)$ and $H_{ff}(x,z)$ to find
\begin{align}
H_{ff}(y,z) &= G_f(y,z)-\frac{\frac{n\delta t}{2} G_f(y,x)G_f(x,z)D(x)}{1+\frac{n\delta t}{2} G_f(x,x)D(x)} \nonumber\\
&=G_f(y,z) - \frac{G_f(y,x)G_f(x,z)}{G_f(x,x)}\frac{1}{1+(n\delta t G_f(x,x)D(x)/2)^{-1}}\nonumber \\
&\sim G_f(y,z)-\frac{G_f(y,x)G_f(x,z)}{G_f(x,x)}\left[1- \frac{2}{n\delta t G_f(x,x) D(x)}\right] \qquad \textrm{ as } n\to\infty.
\end{align}
We also solve the lower right quadrant in the same manner. For shorthand, let us denote
\begin{align}
 \overline{A}(x) &= \frac{1}{n}\sum A(x,t_j) \\
 \overline{A^2}(x) &= \frac{1}{n}\sum_j {A^2}(x,t_j) \\
 \overline{d^2} &= \frac{1}{n}\sum d_j^2 \\
 A_1(x) &= \frac{n\delta t}{2}\left[G_g(x,x)\frac{\overline{d^2}+\overline{A^2}(x)(\delta t)^2}{2D(x)(\delta t)^2} +\partial_z G_g(x,x)\overline{A}(x) \right] \approx\frac{n\beta\delta t}{2}\left[\frac{\overline{d^2}+\overline{A^2}(x)(\delta t)^2}{2D(x)(\delta t)^2}\right] \\
 A_2(x) &= \frac{n\delta t}{2}\left[  G_g(x,x)\overline{A}(x) + \partial_z G(x,x)D(x)\right] \approx \frac{n\beta\delta t}{2} \overline{A}(x) \\
 A_3(x) &=\frac{n\delta t}{2}\left[\partial_yG_g(x,x)\frac{\overline{d^2}+\overline{A^2}(x)(\delta t)^2}{2D(x)(\delta t)^2} +\partial_y\partial_z G_g(x,x)\overline{A}(x) \right]\approx \frac{n\beta\delta t}{2\gamma} \overline{A}(x) \\
 A_4(x) &=\frac{n\delta t}{2}\left[  \partial_y G_g(x,x)\overline{A}(x) + \partial_y\partial_z G(x,x)D(x)\right] \approx \frac{n\beta\delta t}{2\gamma} D(x).
 \end{align}
Where in the approximations we have assumed that $x$ is sufficiently far from $x=0$ so that the boundary condition of the Greens function is insignificant ($\exp(-2x^2/\gamma)\ll 1$). This simplification implies that $G(x,x)=\beta$, $\partial_y G(x,x)=\partial_z G(x,x)=0$, and $\partial^2_{yz}G(x,x)=\beta/\gamma$.
 Eqs.~\ref{eq:Lambda1} and~\ref{eq:Lambda2}, written in terms of these expressions, is
 \begin{align}
 H_{gg}(x,z) &= G_g(x,z) - A_1(x)H_{gg}(x,z)-A_2(x)\partial_yH_{gg}(x,z) \\
 \partial_yH_{gg}(x,z)&=\partial_yG_g(x,z) - A_3(x)H_{gg}(x,z)-A_4(x)\partial_yH_{gg}(x,z).
 \end{align}
We solve this system of intermediate linear equations to obtain
\begin{align}
H_{gg}(x,z) &= \frac{G_g(x,z)(1+A_4(x))-A_2(x)\partial_yG_g(x,z)}{(1+A_1(x))(1+A_4(x))-A_2(x)A_3(x)} \label{eq:Hggxz}\\
 \partial_yH_{gg}(x,z)&=\frac{\partial_yG_g(x,z)}{1+A_4(x)}-\frac{A_3(x)}{1+A_4(x)}\left[\frac{G_g(x,z)(1+A_4(x))-A_2(x)\partial_yG_g(x,z)}{(1+A_1(x))(1+A_4(x))-A_2(x)A_3(x)}\right].\label{eq:yHggxz}
\end{align}
Using Eq.~\ref{eq:Hggxz} and~\ref{eq:yHggxz} we may compute the desired quantity
\begin{align}
H_{gg}(y,z) &= G_g(y,z) - \frac{n\delta t}{2}\left[G_g(y,x)\frac{\overline{d^2}+\overline{A^2}(x)(\delta t)^2}{2D(x)(\delta t)^2} +\partial_zG(y,x)\overline{A}(x) \right]H_{gg}(x,z) \nonumber\\
&\qquad-\frac{n\delta t}{2}\left[ G_g(y,x)\overline{A}(x)+\partial_zG(y,x)D(x) \right]\partial_y H_{gg}(x,z).
\end{align}
Now we may compute the posterior variance in $f$, first by computing
\begin{align}
\lefteqn{\Sigma_{ff}(x,z) = G_f(x,z)-\frac{n\delta t}{2}G_f(x,x)D(x)\Sigma_{ff}(x,z)} \nonumber\\
&+\left(\frac{n\delta t}{2}\right)^2G_f(x,x)\Big[ \overline{A}(x)\overline{A}(x)H_{gg}(x,x)+D^2(x)\partial^2_{yz}H_{gg}(x,x)\Big]\Sigma_{ff}(x,z).
\end{align}
It is now straightforward to find that
\begin{align}
\Sigma_{ff}(x,z) &\approx G_f(x,z)\Bigg\{1+\frac{n\delta t}{2}\beta D(x)-\left(\frac{n\delta t}{2}\right)^2\beta\Big[ \overline{A}(x)\overline{A}(x)H_{gg}(x,x)+D^2(x)\partial^2_{yz}H_{gg}(x,x)\Big] \Bigg\}^{-1}.
\end{align}
We may substitute this expression into Eq.~\ref{eq:Sigmaff} to find that
\begin{align}
\Sigma_{ff}(y,z) &= G_f(y,z)-G_f(y,x)\Bigg\{\frac{n\delta t}{2} D(x)-\left(\frac{n\delta t}{2}\right)^2\Bigg[
 \overline{A}(x)\overline{A}(x)H_{gg}(x,x)+D^2(x)\partial^2_{yz}H_{gg}(x,x)\Bigg]\Bigg\}\Sigma_{ff}(x,z).
\end{align}
We wish to find the leading order $n\to\infty$ behavior of $\Sigma_{ff}(y,y)$. We must proceed with some caution though because the terms $A_1,A_2,A_3,A_4$ embedded in $H_{gg}$ are all $\mathcal{O}(n\delta t/2)$, making $H_{gg}$ effectively $\mathcal{O}(2/n\delta t)$.
\begin{align}
\Sigma_{ff}(y,y) \sim G_f(y,y) - \frac{G_f^2(y,x)}{G_f(x,x)}\left[1-\frac{2}{n\beta\delta t(D(x)-Q(x))}\right],\label{eq:Sigmaffyy}
\end{align} 
where
\begin{align}
Q(x) &= \left(\frac{n\delta t}{2}\right)\Bigg[
 \overline{A}^2(x)H_{gg}(x,x) +D^2(x)\partial^2_{yz}H_{gg}(x,x)\Bigg] \nonumber\\
 &= \left(\frac{n\delta t}{2}\right)\frac{\beta(1+A_4(x))(\overline{A}^2(x)+D^2(x)/\gamma)}{(1+A_1(x))(1+A_4(x))-A_2(x)A_3(x) } \nonumber\\
 &\propto \frac{\overline{A}^2(x)+D^2(x)/\gamma}{1+A_1(x)-\frac{A_2(x)A_3(x)}{1+A_4(x)}}
\end{align}
is an $\mathcal{O}(1)$ term with respect to $n$. To minimize Eq.~\ref{eq:Sigmaffyy}, one must minimize $Q(x)$ with respect to pulling. The effect of pulling is encoded in the variable $\overline{A}(x)$, which we decompose as
\begin{equation}
\overline{A}(x) = D(x)[\overline{F}(x)+\overline{F_a}(x)],
\end{equation}
where $F(x)$ is the mean molecular bond force and $F_a$ is the force applied
by the puling apparatus. $Q(x)$ is minimized with $\overline{A}=0$, or when the applied
drift exactly cancels out the bond drift and diffusivity drift.

%%%%%%%%%%%%%%%%%%%%%%%%%%%%%%%%%%%%%%%%%%%%%%%%%%%%%%%%%%%%%%%
\end{document}